\documentclass[a4paper]{article}
\usepackage{authblk}
\usepackage{booktabs}
\usepackage[dvipdfmx]{graphicx} 
\usepackage[top=30truemm,bottom=25truemm,left=25truemm,right=25truemm]{geometry}
\usepackage{dcolumn}
\usepackage{bm}
\usepackage[percent]{overpic}

\usepackage{dsfont}
\usepackage{amsmath}
\usepackage{braket}
\usepackage{amsfonts}
\usepackage{amssymb}
\usepackage{upgreek}
\usepackage{bbold}
\usepackage{color}
\usepackage{hyperref}
\hypersetup{colorlinks=true, citecolor= blue, linkcolor= blue, filecolor= blue, urlcolor= blue}
\usepackage{comment}
\usepackage{multirow}
\usepackage[dvipsnames]{xcolor}
\usepackage{soul}
\usepackage{lineno}
\usepackage[numbers, sort]{natbib}
\usepackage[normalem]{ulem}

\usepackage{orcidlink}

\usepackage{feynmp-auto}
\usepackage{adjustbox}
\usetikzlibrary{decorations.pathmorphing, arrows.meta}
\usepackage{caption}
\usepackage{subcaption}
\usepackage[T1]{fontenc}

\makeatletter
\patchcmd{\NAT@citex}
  {\@citea\NAT@hyper@\NAT@nmfmt{\NAT@nm}\@extra@b@citeb}
  {\@citea\NAT@nmfmt{\textit{\NAT@nm}}\@extra@b@citeb}
  {}{}
\makeatother

\definecolor{darkgreen}{RGB}{84, 150, 57}

\newcommand{\cbl}{\color{blue}}

\title{Van Vleck 
Excitonic Magnetism in Ruthenium Pyrochlores}
\author[1]{Swetlana Swarup\orcidlink{0000-0001-8651-5387}} 
\affil[1]{School of Physics and Astronomy, University of Minnesota, Minneapolis, Minnesota 55455, USA}

\author[1]{Yang Yang\orcidlink{0000-0001-6375-2005}} 

\author[1,*]{Natalia B. Perkins\orcidlink{0000-0002-9033-1860}}
\affil[*]{Corresponding authors Email: nperkins@umn.edu}
\date{} 

\begin{document}
\maketitle

\begin{abstract}
Strong spin-orbit coupling in $d^4$ systems is expected to stabilize a nonmagnetic $J=0$ singlet ground state, yet many  ruthenium pyrochlores exhibit robust long-range magnetic order. Motivated by this apparent contradiction, we develop a microscopic theory of Van Vleck excitonic magnetism on the pyrochlore lattice. Starting from a multi-orbital Hubbard model with spin-orbit coupling, we derive the effective superexchange interactions within the low-energy singlet--triplet manifold of Ru$^{4+}$ ions. 
We analyze the resulting excitonic Hamiltonian using both the spectrum of triplon excitations and a variational treatment of the condensed phase.
We identify the instability of the nonmagnetic singlet state toward triplon condensation and determine the resulting magnetic phase diagram as a function of the microscopic hopping parameters. The phase diagram reproduces the magnetic orders known from conventional pyrochlore models while also predicting an additional magnetic phase unique to the singlet--triplet description.
Finally, we apply the theory to the pyrochlore ruthenates, with particular emphasis on Nd$_2$Ru$_2$O$_7$, and show that it lies in close proximity to the excitonic quantum critical point. Our results establish a microscopic framework for understanding excitonic magnetism in pyrochlore ruthenates and their magnetic excitation spectrum, providing direct connections to spectroscopic probes,  including Raman scattering.
\end{abstract}

\maketitle


\section*{Introduction}\label{sec:intro}

The interplay of spin--orbit coupling and electronic correlations in transition-metal oxides can give rise to unconventional forms of magnetism that go beyond the standard paradigm of localized moments.  A representative example is provided by $d^4$ systems, where spin--orbit coupling favors a nonmagnetic $J=0$ ground state. At first sight, such systems appear incompatible with magnetic order, since well-defined local moments are absent. Remarkably, however, exchange interactions can drive the coherent condensation of low-energy singlet--triplet excitations. This mechanism, proposed by Khaliullin~\cite{khaliullin2013}, established the concept of Van Vleck \emph{excitonic} magnetism, in which magnetic order arises from the condensation of spin--orbit excitons rather than from preformed local moments.

The excitonic scenario has stimulated extensive theoretical and experimental studies of $4d^4$ ruthenates~\cite{Akbari2014,Souliou2017,Jain2017,Gretarsson2019,chaloupka2019highly,Takahashi2021,singlet2026}, particularly Ca$_2$RuO$_4$~\cite{Akbari2014,KhaliullinPRL2016,Souliou2017,Jain2017,Gretarsson2019}, where excitonic magnetism and its collective excitations have been investigated in considerable detail. It has also inspired extensive work on $5d^4$ iridates~\cite{Cao2014,LagunaPRB2020,ChenPRB2017,DeyPRB2016,Schnait2022,KuschPRB2018,PajskrPRB2016,WilsonPRB2022,AczelPRR2022}, particularly the double perovskites Sr$_2$YIrO$_6$~\cite{Cao2014,LagunaPRB2020} and Ba$_2$YIrO$_6$~\cite{ChenPRB2017,DeyPRB2016,Schnait2022}, where the realization of a $J=0$ ground state and the extent to which excitonic magnetism is realized remain subjects of active debate. Together, these studies have established excitonic magnetism as a realistic mechanism of magnetic order and identified a growing family of candidate materials.

An important open question is how excitonic magnetism is modified by lattice geometry, particularly in frustrated three-dimensional systems. Pyrochlore ruthenates $A_2$Ru$_2$O$_7$, where $A$ is a rare-earth ion, provide an ideal platform to address this question~\cite{Taira1999,Ito2001,Kennedy1995,Ku2018,Laurita2019,wulferding2023,Lee2023,pyrochloreDFT2024}. 
In these materials, Ru ($4d^4$) ions form a network of corner-sharing tetrahedra that combines strong spin--orbit coupling, geometric frustration, and bond-dependent hopping processes. Moreover, the rare-earth series provides a natural route for tuning structural distortions and exchange interactions, enabling a systematic exploration of the excitonic regime. These characteristics make pyrochlore ruthenates a unique setting in which to investigate Van Vleck excitonic magnetism beyond the layered and perovskite compounds studied so far.

Although the local electronic structure of Ru ions suggests a nonmagnetic $J=0$ singlet ground state, many pyrochlore ruthenates develop long-range magnetic order of the Ru sublattice at temperatures of order $100$~K~\cite{Kmiec2006,Zouari2009,Ku2018,Taira1999,Perez2015,Gurgul2007,Chang2010,Bustos2024}. This coexistence of a nominally nonmagnetic ion with robust magnetic order presents a central puzzle for the entire family of compounds. One possible interpretation is that crystal-field effects compete with spin--orbit coupling, driving the system away from the ideal $J=0$ limit toward a more conventional $S=1$ local-moment description, as assumed in previous studies of spin-1 pyrochlore antiferromagnets~\cite{GangChen2018}. An alternative explanation is provided by the excitonic scenario, in which intersite exchange interactions overcome the singlet--triplet gap and drive the condensation of spin--orbit excitons. Establishing whether this mechanism is realized in pyrochlore ruthenates requires a microscopic theory that connects the multi-orbital electronic structure of Ru$^{4+}$ ions to the magnetic ground state and its collective excitation spectrum.

In this work, we develop a microscopic theory of Van Vleck excitonic magnetism on the pyrochlore lattice. Starting from a multi-orbital Hubbard model with spin--orbit coupling, we derive the effective superexchange interactions within the local singlet--triplet manifold of Ru$^{4+}$ ions. We determine the conditions under which the nonmagnetic $J=0$ state becomes unstable toward triplon condensation and construct the resulting magnetic phase diagram. Besides reproducing the magnetic orders familiar from conventional spin-1 pyrochlore models, the excitonic framework predicts an additional magnetic phase that is unique to the singlet--triplet description.

Among the pyrochlore ruthenates, Nd$_2$Ru$_2$O$_7$ provides an ideal testing ground for this theory. Its Ru magnetic order has been characterized by neutron diffraction and magnetization measurements~\cite{Ito2001,Ku2018}, while recent Raman scattering experiments have revealed well-defined low-energy magnetic excitations associated with the Ru sublattice~\cite{wulferding2023,Lee2023}. These complementary experimental results provide strong constraints for a microscopic theory of excitonic magnetism. We therefore apply our microscopic framework to this material and show that it lies in close proximity to the excitonic quantum critical point, providing a unified description of its magnetic ground state and excitation spectrum.

\section*{Results}\label{sec:model}

\subsection*{Lattice and single-ion physics}\label{subsec:localgeometry}

In pyrochlore ruthenates $A_2$Ru$_2$O$_7$, the Ru ions form a network of corner-sharing tetrahedra [see Fig.~\ref{fig:lattice}(a)], which provides the lattice geometry for the excitonic degrees of freedom considered below.The rare-earth $A$-site ions may also carry magnetic moments \cite{Taira1999,Ito2001}. However, their ordering typically occurs at much lower energy scales and can often be treated separately from the Ru-sublattice magnetism. For example, in Nd$_2$Ru$_2$O$_7$, the Nd moments order into an AIAO configuration only below $\sim 1.8$~K \cite{Ku2018}.

On the pyrochlore lattice of Ru ions, each primitive unit cell contains four sublattice sites forming a tetrahedron. Nearest-neighbor bonds are classified by their orientation with respect to the global cubic axes: bonds perpendicular to the $X$, $Y$, and $Z$ directions are referred to as $X(X')$-, $Y(Y')$-, and $Z(Z')$-bonds, respectively. Each Ru ion is surrounded by an oxygen octahedron, and neighboring octahedra share corners, giving rise to bond-dependent local environments. The four sublattice sites thus have distinct local coordinate frames, as illustrated for sites 1 and 4 on the $Z$-bond in Fig.~\ref{fig:lattice}(a).

\begin{figure}[h!]
\centering
\includegraphics[width=\linewidth]{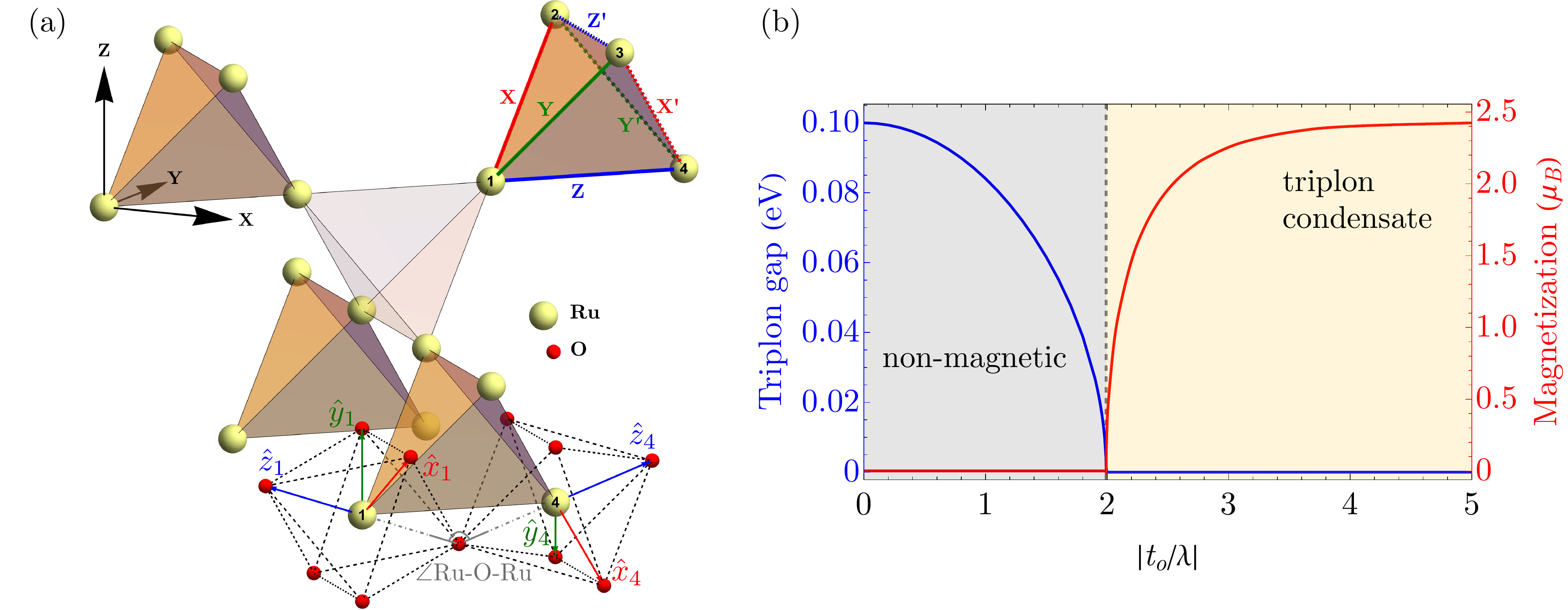}
\caption{\label{fig:lattice}
(a) Pyrochlore lattice of Ru ions, consisting of two types of tetrahedra (dark- and light-shaded). Black arrows $(X,Y,Z)$ indicate the global reference frame. The local reference frames $(\hat{x}_1,\hat{y}_1,\hat{z}_1)$ and $(\hat{x}_4,\hat{y}_4,\hat{z}_4)$, defined by the surrounding oxygen octahedral cages, are shown for sublattice sites 1 and 4, respectively. The nearest-neighbor bonds are marked in red, blue, and green within the top-right tetrahedron. In the ideal pyrochlore structure, the bond angle at the bridging oxygen between two neighboring Ru ions is $\angle\text{Ru-O-Ru}=2\arctan(2\sqrt{2})\approx 141.058^\circ$.
 (b)
 Triplon gap (blue) and the magnetization (red) as a function of oxygen-assisted hopping to spin-orbit coupling constant, $|t_{o}/\lambda|$,  in the ideal structure of pyrochlore lattice. The dashed line marks the phase transition from a non-magnetic singlet ground state to a triplon condensation with finite magnetization, signaling magnetic ordering of the Ru moments. }
\label{fig:lattice}
\end{figure}

The corner-sharing geometry can be understood by starting from two edge-sharing oxygen octahedra aligned with the global coordinate frame and counter-rotating them about the global $[1\bar{1}0]$ axis until one pair of oxygen ions coincides, thereby forming a corner-sharing configuration [see Fig.~\ref{fig:lattice}(a)]. For ideal octahedra, this rotation angle is $\arctan(2\sqrt{2})$, resulting in a bond angle $\angle\text{Ru-O-Ru}=2\arctan(2\sqrt{2})\approx 141.058^\circ$ at the bridging oxygen. As a result, each Ru site acquires its own local Cartesian frame, e.g., $(\hat{x}_1,\hat{y}_1,\hat{z}_1)$ and $(\hat{x}_4,\hat{y}_4,\hat{z}_4)$ in Fig.~\ref{fig:lattice}(a), in which the $t_{2g}$ orbitals are naturally defined. The explicit local axes and the corresponding rotations to the global cubic frame are given in Methods.

At the single-ion level, the local electronic structure is governed by the combined effects of the octahedral crystal field, Hund’s coupling $J_H$, and spin--orbit coupling (SOC) $\lambda$. The crystal field splits the five $d$ orbitals into lower-energy $t_{2g}$ and higher-energy $e_g$ states. Since the $e_g$ states lie far above the energy scales relevant here, we focus exclusively on the $t_{2g}$ manifold. Within the $t_{2g}$ manifold, Hund’s coupling favors states with spin $S=1$ and effective orbital angular momentum $L_{\text{eff}}=1$. SOC further lifts this degeneracy, yielding a $J=0$ singlet ground state and excited $J=1$ and $J=2$ multiplets, with a singlet-triplet gap of $\lambda$ and a triplet-quintet gap of $2\lambda$ \cite{AbragamBleaney1970}. 

To describe the relevant physics, we write the
single-ion Hamiltonian on the $t_{2g}$ manifold as
\begin{align}\label{Ham}
\mathcal{H}_{\mathrm{ion}}=\mathcal{H}_{\text{int}}+\mathcal{H}_{\text{SOC}}+\mathcal{H}_{\Delta},
\end{align}
where $\mathcal{H}_{\text{int}}$ is the multi-orbital Hubbard Hamiltonian that accounts for on-site Coulomb and Hund’s interactions, $\mathcal{H}_{\text{SOC}}$ for the SOC, and $\mathcal{H}_{\Delta}$ for trigonal crystal-field distortions that vanish in the ideal octahedral limit. The SOC term is given by
\begin{equation}\label{eqn:SOC}
\mathcal{H}_{\text{SOC}}=\lambda\sum_i \mathbf{S}_i \cdot \mathbf{L}_i,
\end{equation}
where $\mathbf{S}_i$ and $\mathbf{L}_i$ are the spin and effective orbital angular momentum operators projected onto the $t_{2g}$ manifold, with $L_{\mathrm{eff}}=1$. This coupling splits the local Hilbert space into multiplets of total angular momentum $J$. In the following, we focus on the low-energy sector spanned by the $J=0$ singlet and $J=1$ triplet states. The interaction term reads
\begin{align}
\mathcal{H}_{\text{int}}=\sum_{i}&\bigg(U_1\sum_\alpha n_{i\alpha\uparrow}n_{i\alpha\downarrow}+\frac{1}{2}(U_2-J_H)\sum_{\alpha\neq\alpha',\sigma}n_{i\alpha\sigma}n_{i\alpha'\sigma}+U_2\sum_{\alpha\neq\alpha'}n_{i\alpha\uparrow}n_{i\alpha'\downarrow}\nonumber\\+&J_H\sum_{\alpha\neq\alpha'}d_{i\alpha\uparrow}^\dagger 
d_{i\alpha\downarrow}^\dagger d_{i\alpha'\downarrow}d_{i\alpha'\uparrow}- J_H\sum_{\alpha\neq\alpha'}d_{i\alpha\uparrow}^\dagger d_{i\alpha\downarrow}d_{i\alpha'\downarrow}^\dagger 
d_{i\alpha'\uparrow}\bigg),
\end{align}
where  $d^\dagger_{i\alpha\sigma}$ ($d_{i\alpha\sigma}$) creates (annihilates) an electron at site $i$ in orbital $\alpha$ with spin $\sigma$, $\alpha=yz,xz,xy$ labels the three $t_{2g}$ orbitals, $U_1$ and $U_2$ are the intra- and inter-orbital Coulomb repulsions, $J_H$ is Hund's coupling,
and cubic symmetry enforces $U_1=U_2+2J_H$.

Restricting to the $J=0$ and $J=1$ manifold, the singlet and triplet states can be written in terms of $|L_zS_z\rangle$ eigenstates as
\begin{equation}
    \begin{aligned}
        &\ket{s}=\ket{J=0,J_z=0}=\dfrac{1}{\sqrt{3}}\bigg(\ket{\bar{1}1}-\ket{00}-\ket{1\bar{1}}\bigg),\\
        &\ket{T_1}=\ket{J=1,J_z=1}=\dfrac{1}{\sqrt{2}}\bigg(\ket{\bar{1}0}-\ket{0\bar{1}}\bigg),\\
        &\ket{T_0}=\ket{J=1,J_z=0}=\dfrac{1}{\sqrt{2}}\bigg(\ket{1\bar{1}}+\ket{\bar{1}1}\bigg),\\
        &\ket{T_{\bar{1}}}=\ket{J=1,J_z=-1}=\frac{1}{\sqrt{2}}\bigg(\ket{10}+\ket{01}\bigg).
    \end{aligned}
\end{equation}
The triplet states $\{\ket{T_1},\ket{T_0},\ket{T_{\bar{1}}}\}$ are eigenstates of $J_z$ and form the magnetic basis. However, in the absence of hopping the ground state is a $J=0$ singlet, invariant under time reversal. It is therefore more natural to work in a basis that respects this symmetry. We introduce the Cartesian “triplon” basis $\{\ket{T^x},\ket{T^y},\ket{T^z}\}$ \cite{remund2022semi}:
\begin{align}
        \ket{T^x}&=\dfrac{1}{i\sqrt{2}}(\ket{T_1}-\ket{T_{\bar{1}}}),\nonumber\\
        \ket{T^y}&=\dfrac{1}{\sqrt{2}}(\ket{T_1}+\ket{T_{\bar{1}}}),\label{jmbasis}\\
        \ket{T^z}&=i\ket{T_0},\nonumber
\end{align}
which is manifestly time-reversal invariant. Together with the singlet state, these states define the local low-energy Hilbert space on each site $i$:
\begin{align}
    \{\ket{\tau_i^\alpha}\}=\{\ket{s_i},\ket{T^x_i},\ket{T_i^y},\ket{T_i^z}\}\label{eqn:taus}
\end{align} enumerated by $\alpha=0,1,2,3$, where $\alpha=0$ labels the singlet state and $\alpha=1,2,3$ label three triplet states. This construction is closely related to the excitonic representation introduced by Khaliullin~\cite{khaliullin2013} and further developed for tetragonal ruthenates such as Ca$_2$RuO$_4$~\cite{Akbari2014,Jain2017}, where a strong tetragonal crystal field splits the triplet manifold and effectively selects a reduced set of states, allowing one to define a pseudospin-1 basis in terms of $\{ \ket{s_i}, \ket{T^x_i}, \ket{T^y_i} \}$.

In contrast, for pyrochlore ruthenates the RuO$_6$ octahedra are uniformly trigonally compressed. Although the trigonal distortion has the same magnitude for every octahedron, the octahedra themselves are differently oriented within the pyrochlore lattice. Consequently, the compression axis coincides with 
the local $[111]$ direction of each Ru site, described by the unit vector
$\mathbf{e}_{i,[111]}$ on site $i$. Throughout this work we adopt the convention $\Delta>0$, corresponding to trigonal compression experimentally observed in the A$_2$Ru${}_2$O${}_7$ family of compounds~\cite{Kennedy1995,Laurita2019,Ku2018,Pawar2017,Chakraborty2026,Chatterjee2024,Castro2021,Kennedy1996,Yamamoto1994,Museur2026,Taira2003,Bustos2024}. The resulting trigonal crystal field is described by{
\begin{align}\label{eq:trigonal}
    \mathcal{H}_\Delta=
    \sum_i\Delta     \left(\mathbf{L}_i\cdot\mathbf{e}_{i,[111]}\right)^2,
\end{align}
}which splits the cubic $t_{2g}$ manifold into a lower $e_g'$ doublet and an upper $a_{1g}$ singlet. In the limit of a very large trigonal splitting, the lower $e_g'$ doublet becomes completely filled, resulting in the nonmagnetic configuration $e_g'^4a_{1g}^0$ with $S=L=0$. In pyrochlore ruthenates, however, the trigonal crystal-field splitting is smaller than the Hund's coupling, making it energetically favorable to promote one electron into the $a_{1g}$ orbital. The resulting high-spin configuration, $e_g'^3a_{1g}^1$, has $S=1$ and $L_{\rm eff}=1$, which are subsequently coupled by the spin-orbit interaction into the $J=0$, $J=1$, and $J=2$ multiplets discussed above.

\subsection*{Effective low-energy Hamiltonian from superexchange}\label{subsec:superexchange}

We next derive the superexchange interaction between the local singlet-triplet degrees of freedom. Electron hopping between neighboring Ru ions generates virtual charge fluctuations out of the $d^4_i d^4_j$ manifold into intermediate $d^3_i d^5_j$ ($d^5_i d^3_j$) configurations. 
Integrating out these virtual states to second order in the hopping and projecting back onto the local $J=0,1$ manifold yields an effective superexchange Hamiltonian for the excitonic degrees of freedom.
The hopping Hamiltonian is
\begin{equation}\label{eq:hopping}
\mathcal{H}_{\text{t}}
=
\sum_{i,j}\sum_{\alpha,\beta}\sum_{\sigma,\sigma'}
t_{ij}^{\alpha\sigma,\beta\sigma'}
d^{\dagger}_{i\alpha\sigma}d_{j\beta\sigma'} .
\end{equation}
Here $d_{i\alpha\sigma}$ annihilates an electron on site $i$ in the local $t_{2g}$ orbital 
$\alpha=yz,xz,xy$ and local spin state $\sigma$. 
The hopping amplitudes are bond-dependent and are defined in the local orbital--spin basis. Their matrix structure reflects the geometry of the corner-sharing oxygen octahedra and the corresponding overlap of $t_{2g}$ orbitals in the local coordinate frames. 

As an example, the direct hopping matrix for the $Z$-bond between sites 1 and 4 is shown in Table~\ref{tbl:hop}. The hopping on other bonds follows from the $C_3$ symmetry of the cubic point group $T_d$. The derivation of the hopping parameters in terms of Slater--Koster parameters is given in Supplementary Sec.~\ref{slaterkoster}. In the ideal structure of pyrochlore considering only the oxygen-mediated hopping, we have
\begin{align}\label{t1t2-to}
    t_1 = t_o/9,\quad t_2=-8t_o/9,\quad t_3=0,\quad t_4=0,
\end{align}
where the oxygen-mediated hopping amplitude is $t_{o}\equiv pd\pi^2/\Delta_{\mathrm{pd}}$ determined by the overlap $pd\pi$ between $t_{2g}$ orbital and the oxygen $p$ orbital and the charge transfer $\Delta_{\mathrm{pd}}$.  

\begin{table}[!h]
        \centering
        \begin{tabular}{|c|c|c|c|}
            \hline
                & $d_{4,yz}$ & $d_{4,xz}$ & $d_{4,xy}$ \\\hline
            $d_{1,yz}$ & $t_1$ & $t_2$ & $t_4$ \\\hline
            $d_{1,xz}$ & $t_2$ & $t_1$ & $t_4$ \\\hline
            $d_{1,xy}$ & $-t_4$ & $-t_4$ & $t_3$  \\\hline
        \end{tabular} \quad
        {\renewcommand{\arraystretch}{1.33}
        \begin{tabular}{|c|c|c|}
        \hline
        &$\ket{\uparrow}_4$ & $\ket{\downarrow}_4$ \\
        \hline
    $\ket{\uparrow}_1$ & $\frac{1}{3}$ & $-\frac{2}{3}-\frac{2 i}{3}$ \\ \hline
     $\ket{\downarrow}_1$ & $\frac{2}{3}-\frac{2 i}{3}$ & $\frac{1}{3}$ \\ \hline
    \end{tabular}
    \caption{\label{tbl:hop} (left) Direct hopping amplitudes from site 1 to 4 between $t_{2g}$ orbitals expressed in their respective local basis on the $Z$-bond. (right) Spin states expressed in the local basis. }}
    \end{table}

Formally, the superexchange Hamiltonian is obtained by second-order degenerate perturbation theory with the hopping Hamiltonian. 
We start from the two-site $d_i^4d_j^4$ low-energy manifold, in which each Ru ion has $L_{\mathrm{eff}}=1$ and $S=1$.  
Now we construct the second-order degenerate perturbation
\begin{equation}\label{eqn:2pt}
\sum_{ll'}\sum_I
\frac{
\langle l'|\mathcal{H}_t|I\rangle
\langle I|\mathcal{H}_t|l\rangle
}
{E_0-E_I} |l'\rangle\langle l|,
\end{equation}
where $\ket{l}$ and $\ket{l'}$ belong to the initial $d_i^4d_j^4$ manifold with energy $E_0$, while $\ket{I}$ denotes intermediate charge-transfer states of the form $d_i^3d_j^5$ ($d_i^5d_j^3$). 

Each contribution in Eq.~\eqref{eqn:2pt} contains two hopping events: the first creates a virtual $d_i^3d_j^5$ ($d_i^5d_j^3$) charge excitation, and the second returns the system to the $d_i^4d_j^4$ manifold. Expressing Eq.~\eqref{eqn:2pt} in terms of spin-$1$ operators and $L=1$ orbital angular momentum operators yields the Kugel--Khomskii Hamiltonian~\cite{Kugel1982}. 

Here we are instead interested in the singlet–triplet subspace. Projecting Eq.~\eqref{eqn:2pt} onto this subspace, we obtain a bond-dependent exchange Hamiltonian for the excitonic degrees of freedom. For convenience, we introduce hard-core bosons $\tau_{i\alpha}$ defined in Eq.~\eqref{eqn:taus} subject to the single-occupancy constraint $s_i^\dagger s_i+T^{x\dagger}_i T^x_i+T^{y\dagger}_i T^y_i+T^{z\dagger}_i T^z_i=1$. In terms of these operators, the resulting nearest-neighbor interaction takes the form
\begin{equation}\label{eq:Heff}
\mathcal{H}_{\mathrm{eff}}^{(4)}
=
\sum_{\langle ij\rangle}
\sum_{\alpha\alpha'\beta\beta'}
J^{\alpha\alpha'\beta\beta'}_{ij}\,
\tau_{i\alpha}^\dagger \tau_{j\alpha'}^\dagger
\tau_{i\beta}\tau_{j\beta'}
+ \mathrm{h.c.},
\end{equation}
where $J^{\alpha\alpha'\beta\beta'}_{ij}$ encodes the bond-dependent exchange interactions generated by the virtual charge-transfer processes. The explicit construction of the projected bond Hamiltonians and the generation of symmetry-related bonds are described in the Methods. 
The projected quartic superexchange Hamiltonian \eqref{eq:Heff} retains the full microscopic details of $\mathcal{H}_{\mathrm{int}}$. Similarly, $\mathcal{H}_{\mathrm{SOC}}$ and $\mathcal{H}_{\Delta}$ can be expressed in terms of the hard-core boson operators. Together with $\mathcal{H}_{\mathrm{eff}}^{(4)}$, they constitute the full low-energy effective Hamiltonian. In the following, for simplicity, we consider the ideal octahedral structure without trigonal distortion, since the trigonal field $\mathcal{H}_\Delta$ enters in a manner analogous to the spin--orbit coupling and does not qualitatively change our results, as will be shown when we specialize to the A${}_2$Ru${}_2$O${}_7$ family of compounds.

\begin{figure}[h]
    \centering
    \includegraphics[width=\linewidth]{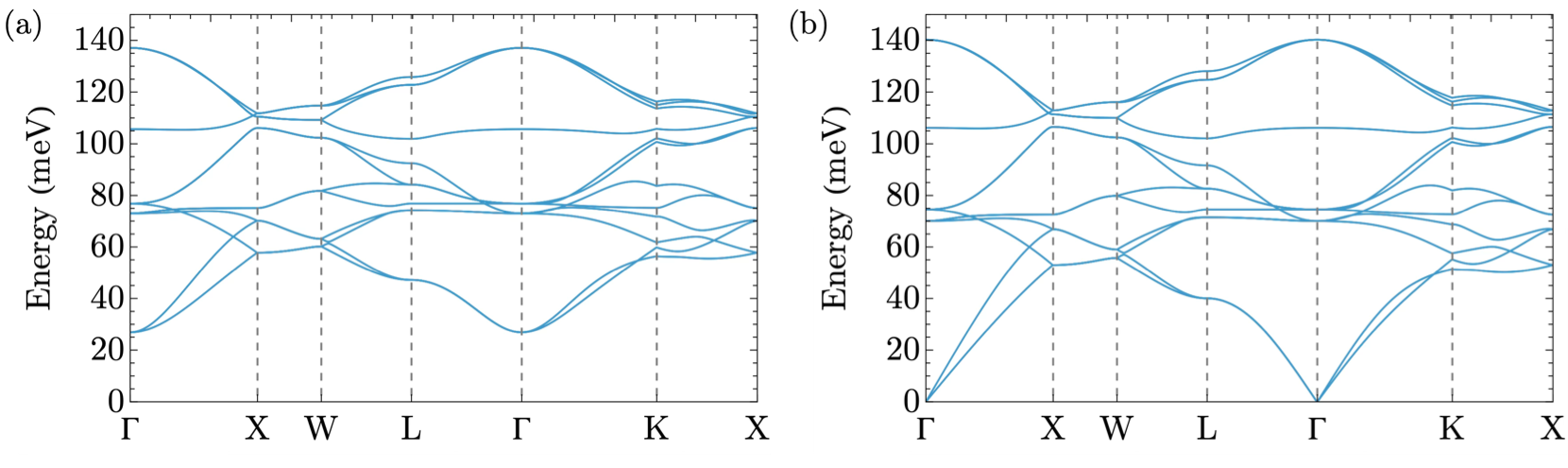}
    \caption{Dispersion of the 12 triplon bands above the singlet ground state along the high-symmetry path in momentum space, computed with $U_2=1.7$~eV, $J_H=0.35$~eV, and $\lambda=0.1$~eV: (a) below the critical hopping at $|t_{o}/\lambda|=1.9$ and (b) at the critical hopping $|t_{o}/\lambda|=1.9923$. 
    }
    \label{fig:gapdisp}
\end{figure}

\subsection*{Magnetic instability of the singlet ground state}\label{subsec:spectrum}
The single-ion singlet ground state favored by spin--orbit coupling becomes unstable in the presence of superexchange. To demonstrate this, we assume a macroscopic singlet condensate $s\approx s^\dagger$ and replace the singlet operators $s$ and $s^\dagger$ with $\sqrt{1-T^{x\dagger} T^x-T^{y\dagger} T^y-T^{z\dagger} T^z}$, keeping terms up to the quadratic order in the triplon operators $T^x,T^y$ and $T^z$. Stability of the singlet ground state requires a finite gap in the triplon excitation spectrum. As the hopping strength, hence the superexchange, increases, this gap eventually closes, marking a quantum phase transition to a magnetically ordered ground state composed of an admixture of singlets and triplets. 

As an example, we present the triplon gap closing for a typical Ru parameter set: $U_2=1.7$~eV, $J_H=0.35$~eV, and $\lambda=0.1$~eV. Fig.~\ref{fig:lattice} (b) shows the triplon excitation gap as a function of the dominant oxygen-mediated hopping $|t_o/\lambda|$. The gap closes rapidly near the critical hopping. The full triplon dispersion at $|t_o/\lambda|=1.9$ [Fig.~\ref{fig:gapdisp} (a)] shows a two-fold degenerate triplon excitation gap around $27$ meV at the $\Gamma$ point. Upon increasing $|t_o/\lambda|$ to the critical value $1.9923$, this gap closes at the $\Gamma$ point [Fig.~\ref{fig:gapdisp} (b)], signaling triplon condensation and the onset of magnetic order on the Ru sublattice.

\subsection*{Triplon condensate}\label{subsec:variational}

When the triplon gap closes at the $\Gamma$  point, the singlet ground state becomes unstable toward triplon condensation. To determine the new ground state and the corresponding magnetic order, we construct a variational wavefunction that preserves the crystallographic unit cell, so that the magnetic unit cell consists of a single tetrahedron. On each tetrahedron, we write a product state over the four sublattices: 
\begin{equation}\label{eq:Psi}
\Ket{\Psi}=\prod_{\mu=1,2,3,4}\bigg(\cos{\theta_\mu}\,\ket{s}+\sin{\theta_\mu \ket{\mathbf{d}_{\mu}}}\bigg),
\end{equation}
with variational parameters $\theta_\mu$ and variational states $|\mathbf{d}_\mu\rangle$, which are normalized states within the $J=1$ manifold, parametrized by complex unit vectors $\mathbf{d}_\mu \equiv(d^x_\mu,d^y_\mu,d^z_\mu)$ such that $\ket{\mathbf{d}_\mu}=\sum_{\alpha=x,y,z} d^{\alpha*}_\mu \ket{T^\alpha}$ . 

The variational parameter $\theta_\mu$ controls the triplon condensate density, while $\mathbf{d}_\mu$ determines how the condensate is distributed among three triplon flavors. Within this ansatz, the singlet and triplon occupations on each sublattice $\mu$ are
\begin{equation}
\braket{n_S}_\mu=\cos^2\theta_\mu,\qquad \sum_{\alpha=x,y,z}\braket{n_{T^\alpha}}_\mu
=
\sin^2\theta_\mu.
\end{equation}
The distribution among the three triplon flavors also determines the direction of the ordered moment. Following Refs.~\cite{khaliullin2013,chaloupka2019highly}, we decompose the magnetic moment operator
\begin{equation}
\mathbf M = 2\mathbf S+\mathbf L
\end{equation}
within the low-energy singlet--triplet manifold into two contributions,
\begin{equation}
\mathbf M = \mathbf M_1+\mathbf M_2.
\end{equation}
The first term is the Van Vleck excitonic magnetism arising from singlet--triplet transitions,
\begin{equation}
\mathbf M_1
=
-i\sqrt{6}
\left(
s^\dagger \mathbf T
-
\mathbf T^\dagger s
\right),
\end{equation}
while the second term describes the intrinsic magnetic moment carried by the spin-1 triplons,
\begin{equation}
\mathbf M_2
=
\frac{1}{2}\mathbf J
=
-\frac{i}{2}
\left(
\mathbf T^\dagger\times\mathbf T
\right).
\end{equation}
Evaluating these operators within the variational state gives
\begin{align}
\braket{\mathbf M_1}_\mu
&=
2\sqrt{6}\,
\sin\theta_\mu\cos\theta_\mu\,
\mathbf v_\mu,\\
\braket{\mathbf M_2}_\mu
&=
2\sin^2\theta_\mu\,
(\mathbf u_\mu\times\mathbf v_\mu),
\end{align}
where $\mathbf{u}_\mu$ and $\mathbf{v}_\mu$ are the real and imaginary parts of the complex unit vector $\mathbf{d}_\mu=\mathbf{u}_\mu + i \mathbf{v}_\mu$. Thus, the order parameter of the excitonic phase is not simply the triplet density, but rather a coherent hybridization amplitude between the singlet and triplet sector.  In the parameter regimes discussed below, the optimized solutions are dominated by $\mathbf{v}_\mu$, which provides magnetic moment $\mathbf{M}_1$ up to $2.45\mu_B$ per Ru ion [Fig.~\ref{fig:lattice} (b)]. 
 The ordered moment is therefore predominantly of Van Vleck character, with only a subleading contribution from the intrinsic triplon moment $\mathbf M_2$.

The new ground state is obtained by minimizing the variational energy 
\begin{equation}\label{eqn:Evar}
\begin{aligned}
E_{\mathrm{var}}
=&
\braket{\Psi|\mathcal H_{\mathrm{eff}}^{(4)}+\mathcal{H}_{\mathrm{SOC}}|\Psi}\\
=&
\braket{\Psi_{14}|\mathcal H^{Z}_{14}|\Psi_{14}}
+
\braket{\Psi_{12}|\mathcal H^{X}_{12}|\Psi_{12}}+
\braket{\Psi_{13}|\mathcal H^{Y}_{13}|\Psi_{13}}\\&+
\braket{\Psi_{32}|\mathcal H^{Z'}_{32}|\Psi_{32}}+
\braket{\Psi_{43}|\mathcal H^{X'}_{43}|\Psi_{43}}
+
\braket{\Psi_{42}|\mathcal H^{Y'}_{42}|\Psi_{42}}
\end{aligned}
\end{equation}
within a single tetrahedron with 28 real parameters. Each sublattice site contributes seven real variational parameters: one mixing angle $\theta_\mu$ and the six components of $\mathbf{u}_\mu$ and $\mathbf{v}_\mu$, subject to the normalization constraint $|\mathbf d_\mu|^2=1$.

\subsection*{Phase diagram}\label{subsec:phase}
\begin{figure}[t]
    \centering
    \includegraphics[width=\linewidth]{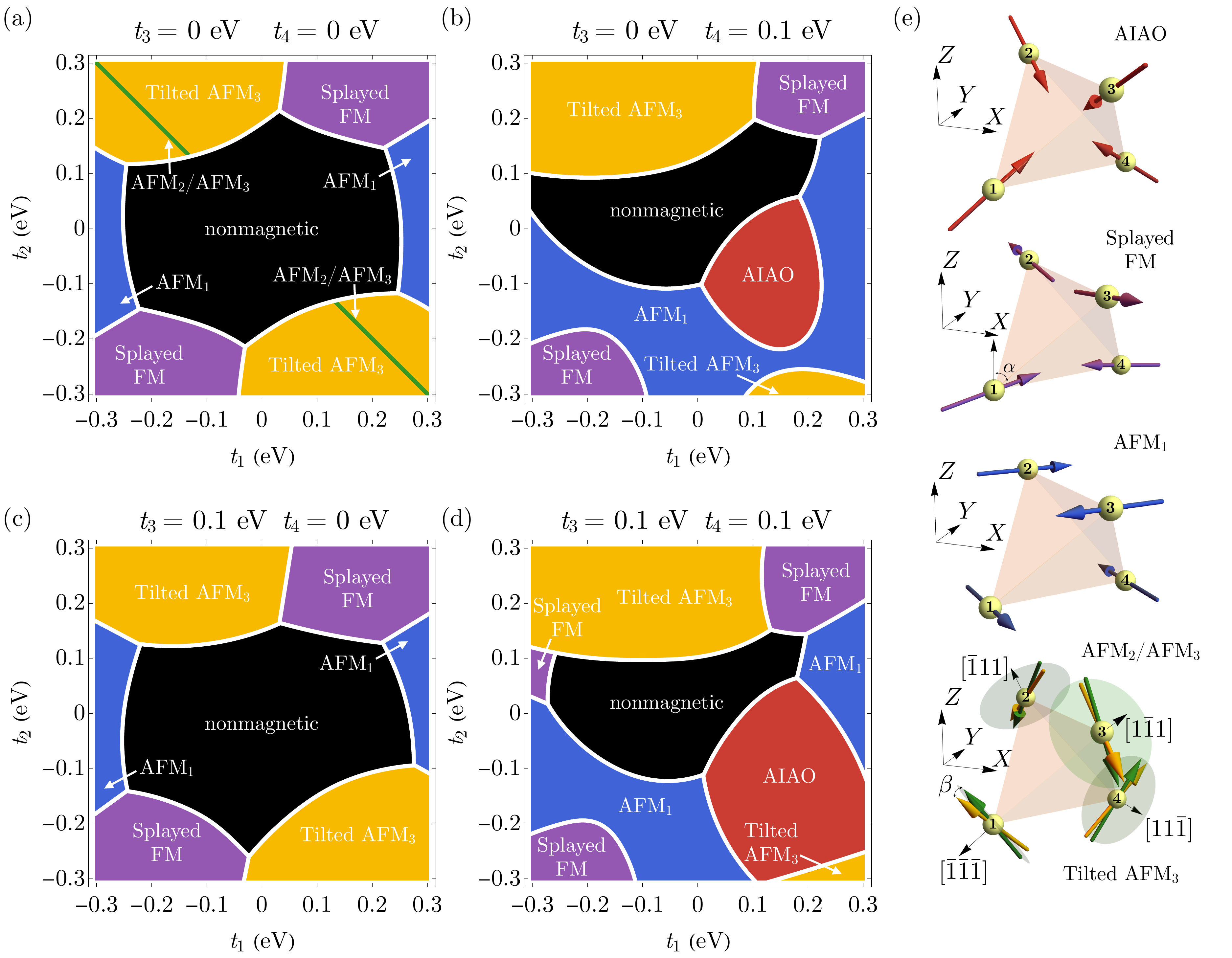}
    \caption{$\mathbf{q}=0$ phase diagram of the microscopic model as a function of the dominant hoppings $t_1$ and $t_2$ obtained by minimizing the variational ground-state energy at (a) $t_3=t_4=0$ eV, (b) $t_3=0$ eV and $t_4=0.1$ eV, (c) $t_3=0.1$ eV and $t_4=0$ eV, (d) $t_3=0.1$ eV and $t_4=0.1$ eV. Along the green line in (a), where $t_1=-t_2$ and $t_3=t_4=0$, an accidental $U(1)$ symmetry emerges that allows all four spins to rotate by a common angle within their respective local planes perpendicular to the axis pointing toward the center of the tetrahedron. The spin configuration of each phase is depicted in (e): the splayed FM is parametrized by the splay angle $\alpha$ measured from the $z$-axis, and the tilted AFM${}_3$ state by the tilt angle $\beta$ measured out of these local planes.}
    \label{fig:phasediagt1t2}
\end{figure}

The ordering of the condensed excitonic states is obtained by minimizing the variational energy, Eq.~\eqref{eqn:Evar}, with respect to $\theta_\mu$, $\mathbf{u}_\mu$, and $\mathbf{v}_\mu$. 
We first focus on the dominant oxygen-mediated hopping channels $t_1$ and $t_2$, Eq.~\eqref{t1t2-to}, which largely determine the magnetic ordering, and map out the resulting $\mathbf{q}=0$ phase diagram in the $(t_1,t_2)$ plane. 
 To illustrate the richness of the excitonic condensate, we examine four representative cuts corresponding to different values of $t_3$ and $t_4$, shown in Fig.~\ref{fig:phasediagt1t2}, while keeping the remaining parameters fixed at $U_2=1.70$~eV, $J_H=0.35$~eV, and $\lambda=0.1$~eV. 
  As shown below, the resulting phase diagram reproduces all of the classical $\mathbf{q}=0$ magnetic orders of the conventional $S=1$ pyrochlore model~\cite{GangChen2018}, while also stabilizing an additional tilted AFM$_3$ phase unique to the excitonic description.

The resulting $\mathbf{q}=0$ phase diagrams are shown in Fig.~\ref{fig:phasediagt1t2}. The different regions correspond to distinct magnetic orderings of the triplon condensate, each specified by four unit vectors, one on each sublattice of a tetrahedron:
\begin{align}
    \text{(i) all-in-all-out (AIAO):}\quad
    &\mathbf{e}_1=\tfrac{1}{\sqrt{3}}(1,1,1),\ \mathbf{e}_2=\tfrac{1}{\sqrt{3}}(1,-1,-1),\nonumber\\
    &\mathbf{e}_3=\tfrac{1}{\sqrt{3}}(-1,1,-1),\ \mathbf{e}_4=\tfrac{1}{\sqrt{3}}(-1,-1,1),\label{eqn:AIAO}\\
    \text{(ii) splayed ferromagnet (splayed FM):}\quad
    &\mathbf{e}_1=\left(\tfrac{\sin\alpha}{\sqrt{2}},\tfrac{\sin\alpha}{\sqrt{2}},\cos\alpha\right),\ \mathbf{e}_2=\left(-\tfrac{\sin\alpha}{\sqrt{2}},\tfrac{\sin\alpha}{\sqrt{2}},\cos\alpha\right),\nonumber\\
    &\mathbf{e}_3=\left(\tfrac{\sin\alpha}{\sqrt{2}},-\tfrac{\sin\alpha}{\sqrt{2}},\cos\alpha\right),\ \mathbf{e}_4=\left(-\tfrac{\sin\alpha}{\sqrt{2}},-\tfrac{\sin\alpha}{\sqrt{2}},\cos\alpha\right),\\
    \text{(iii) coplanar XY antiferromagnet (AFM}_1\text{):}\quad
    &\mathbf{e}_1=\tfrac{1}{\sqrt{2}}(1,-1,0),\ \mathbf{e}_2=\tfrac{1}{\sqrt{2}}(1,1,0),\nonumber\\
    &\mathbf{e}_3=\tfrac{1}{\sqrt{2}}(-1,-1,0),\ \mathbf{e}_4=\tfrac{1}{\sqrt{2}}(-1,1,0),\\
    \text{(iv) coplanar XY antiferromagnet (AFM}_2\text{):}\quad
    &\mathbf{e}_1=\tfrac{1}{\sqrt{2}}(1,-1,0),\ \mathbf{e}_2=\tfrac{1}{\sqrt{2}}(-1,-1,0),\nonumber\\
    &\mathbf{e}_3=\tfrac{1}{\sqrt{2}}(1,1,0),\ \mathbf{e}_4=\tfrac{1}{\sqrt{2}}(-1,1,0),\\
    \text{(v) noncoplanar XY antiferromagnet (AFM}_3\text{):}\quad
    &\mathbf{e}_1=\tfrac{1}{\sqrt{6}}(-1,-1,2),\ \mathbf{e}_2=\tfrac{1}{\sqrt{6}}(-1,1,-2),\nonumber\\
    &\mathbf{e}_3=\tfrac{1}{\sqrt{6}}(1,-1,-2),\ \mathbf{e}_4=\tfrac{1}{\sqrt{6}}(1,1,2),
\end{align}
where $\alpha$ denotes the splay angle measured from the Z-axis. These classical $\mathbf{q}=0$ orderings can also be classified by the irreducible representations of the $O_h$ point group (see Ref.~\cite{Huebsch2022} for details).
The AFM$_1$ and AFM$_2$ orderings are related by reversing the spins on sublattices 2 and 3 and therefore represent distinct magnetic phases.
The AFM$_2$ and AFM$_3$ orderings, by contrast, form a degenerate manifold of states connected by an accidental $U(1)$ symmetry that rotates each spin simultaneously within the local plane perpendicular to axes defined by AIAO states \eqref{eqn:AIAO}.

An extended region of the phase diagram shown in Fig.~\ref{fig:phasediagt1t2} is occupied by a tilted noncoplanar XY antiferromagnetic state. We refer to this state as tilted AFM$_3$ and parametrize it as
\begin{equation}
\begin{aligned}
    &\mathbf{e}_1=\left(-\tfrac{\sin\alpha}{\sqrt{2}},-\tfrac{\sin\alpha}{\sqrt{2}},\cos\alpha\right),\ \mathbf{e}_2=\left(-\tfrac{\sin\alpha}{\sqrt{2}},\tfrac{\sin\alpha}{\sqrt{2}},-\cos\alpha\right),\\
    &\mathbf{e}_3=\left(\tfrac{\sin\alpha}{\sqrt{2}},-\tfrac{\sin\alpha}{\sqrt{2}},-\cos\alpha\right),\ \mathbf{e}_4=\left(\tfrac{\sin\alpha}{\sqrt{2}},\tfrac{\sin\alpha}{\sqrt{2}},\cos\alpha\right),
\end{aligned}
\end{equation}
where $\alpha$ again denotes the splay angle measured from the global $Z$-axis and setting $\alpha = \arctan(1/\sqrt{2})$ recovers the AFM$_3$ state. We can define the tilting angle $\beta \equiv \alpha - \arctan(1/\sqrt{2})$ to quantify the deviation from the degenerate  AFM$_2$/AFM$_3$ manifold. Different from the conventional $S=1$ model, the AFM$_2$/AFM$_3$ degeneracy survives only along the line $t_1 = -t_2$ with $t_3 = t_4 = 0$.

\subsection*{Excitation spectrum of triplon condensate}
Over most of the phase diagram, the ground state develops long-range order through triplon condensation and is described by
 the variational wavefunction of Eq.~(\ref{eq:Psi}) To compute excitations above the ground state with triplon condensate, we introduce a unitary transformation
\begin{align}
\begin{pmatrix}\label{eq:unitary}
       |s\rangle\\
       |T^x\rangle\\
       |T^y\rangle\\
       |T^z\rangle
    \end{pmatrix}=
    \left(
\begin{array}{cccc}
 \cos \theta  & 0 & 0 & i \sin \theta  \\
 (u_x + i v_x) \sin \theta  & -\frac{u_y- i v_y}{\sqrt{u_x^2+u_y^2+v_x^2+v_y^2}} & -\frac{v_x v_z}{\sqrt{u_x^2+u_y^2+v_x^2+v_y^2}} & (-i u_x+v_x) \cos \theta  \\
 (u_y + i v_y) \sin \theta  & \frac{u_x - i v_x}{\sqrt{u_x^2+u_y^2+v_x^2+v_y^2}} & -\frac{v_y v_z}{\sqrt{u_x^2+u_y^2+v_x^2+v_y^2}} & (-i u_y+v_y) \cos \theta  \\
 (u_z + i v_z) \sin \theta  & 0 & \sqrt{u_x^2+u_y^2+v_x^2+v_y^2} & (-i u_z +v_z) \cos \theta \\
\end{array} \right)\begin{pmatrix}
       |\Psi\rangle\\
       |\Phi_1\rangle\\
       |\Phi_2\rangle\\
       |\Phi_3\rangle
    \end{pmatrix}
\end{align}
 that rotates the original basis $\{|s\rangle,|T^x\rangle,|T^y\rangle,|T^z\rangle\}$ into a new orthonormal basis $\{|\Psi\rangle,|\Phi_1\rangle,|\Phi_2\rangle,|\Phi_3\rangle\}$,  where $|\Psi\rangle$ is the variational ground state corresponding to triplon condensate and $|\Phi_1\rangle,|\Phi_2\rangle,|\Phi_3\rangle$ are new excited states.  
 Equation~\eqref{eq:Heff} can then be rewritten in terms of the corresponding hard-core bosons 
  $\{\Psi,\Phi_1,\Phi_2,\Phi_3\}$, 
  subject to the single-occupancy constraint $\Psi^\dagger\Psi+\Phi_1^\dagger\Phi_1+\Phi_2^\dagger\Phi_2+\Phi_3^\dagger \Phi_3=1$. Assuming that the $\Psi$ bosons condense, we approximate $\Psi\approx\Psi^\dagger$ and  replace them with $\sqrt{1-\Phi_1^\dagger\Phi_1-\Phi_2^\dagger\Phi_2-\Phi_3^\dagger \Phi_3}$. Expanding Eq.~\eqref{eq:Heff} to quadratic order in $\Phi_1$, $\Phi_2$, and $\Phi_3$. The resulting quadratic Hamiltonian yields the excitation spectrum of the triplon-condensed phase.

\begin{figure}
    \centering
    \includegraphics[width=\linewidth]{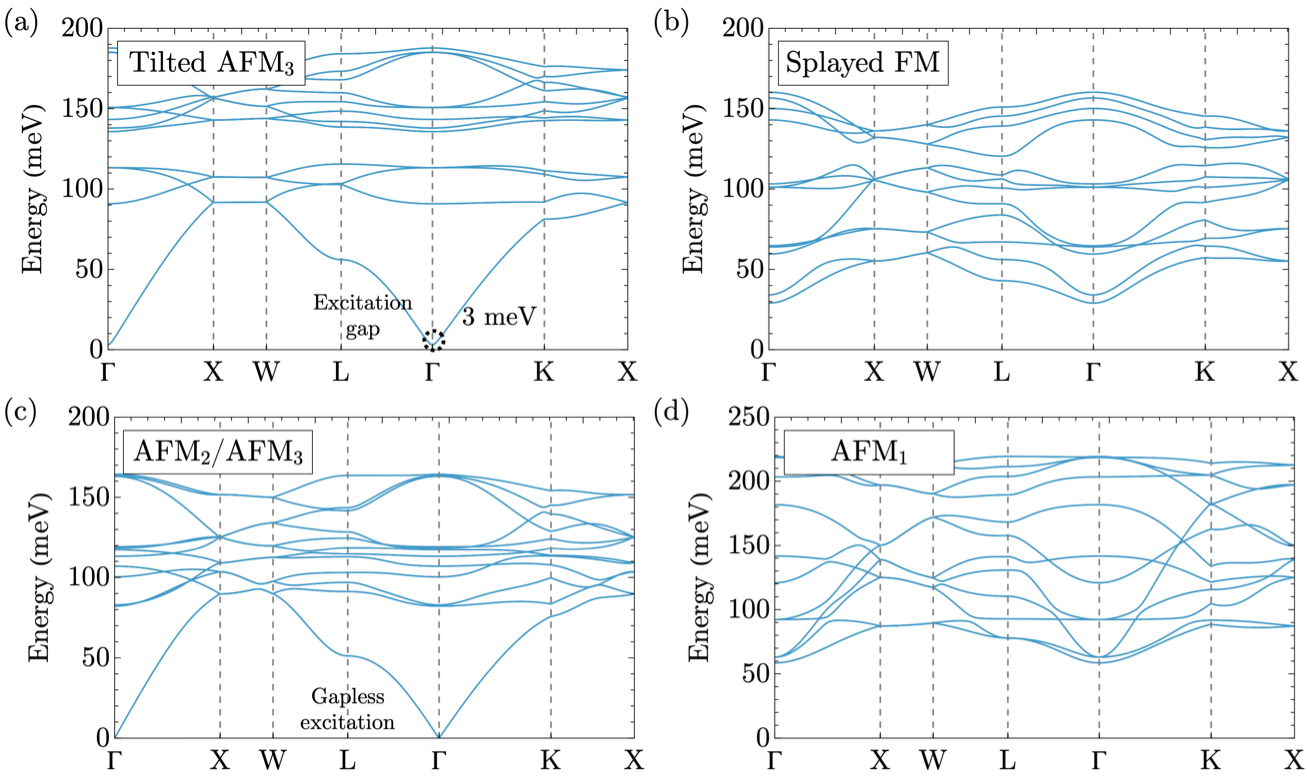}
    \caption{Excitation of triplon condensate with four magnetic sturcture shown in Fig.~\ref{fig:phasediagt1t2}(a), fixing $t_3=t_4=0$. (a) Tilted AFM${}_3$ ($t_1=-0.1$~eV, $t_2=0.2$~eV), (b) Splayed FM ($t_1=0.1$~eV, $t_2=0.2$~eV), (c) AFM${}_2$/AFM${}_3$ ($t_1=-0.15$~eV, $t_2=0.15$~eV), and (d) AFM${}_1$ ($t_1=0.3$~eV, $t_2=0.0$~eV).}
    \label{fig:triplonExcitation}
\end{figure}

We compute the excitation spectrum for the four ordered ground states shown in Fig.~\ref{fig:phasediagt1t2} (a). The results are presented in Fig.~\ref{fig:triplonExcitation}. The tilted AFM${}_3$ state, obtained by a small tilting angle $\beta$ away from the AFM${}_2$/AFM${}_3$ manifold, exhibits a pseudo-Goldstone mode of approximately $3$ meV at the $\Gamma$ point, in contrast to the true Goldstone mode that develops in the AFM${}_2$/AFM${}_3$ state [Fig.~\ref{fig:triplonExcitation}(a) and (c)]. This low-energy pseudo-Goldstone mode provides a clear experimental signature  of the tilted AFM${}_3$ state, whereas the remaining ordered phases exhibit a sizable excitation gap at the $\Gamma$ point [Fig.~\ref{fig:triplonExcitation}(b) and (d)]. 

\subsection*{Application to A${}_2$Ru${}_2$O${}_7$}

We now apply the Van Vleck magnetism framework to ruthenate pyrochlores A${}_2$Ru${}_2$O${}_7$. The experimentally reported properties of the Ru subsystem in A${}_2$Ru${}_2$O${}_7$  are summarized in Table~\ref{tab:A2Ru2O7}. Across this family of compounds, the Ru moments order at temperatures between $80$~K and $150$~K. Structurally, the family of A${}_2$Ru${}_2$O${}_7$ compounds all deviate from the ideal pyrochlore geometry, with the bond angle $\angle\text{Ru-O-Ru}$ differing significantly from $141.058^\circ$. Therefore, we also include the trigonal distortion in our calculation to properly account for magnetic properties in real materials.

\begin{figure}
    \centering
    \includegraphics[width=\linewidth]{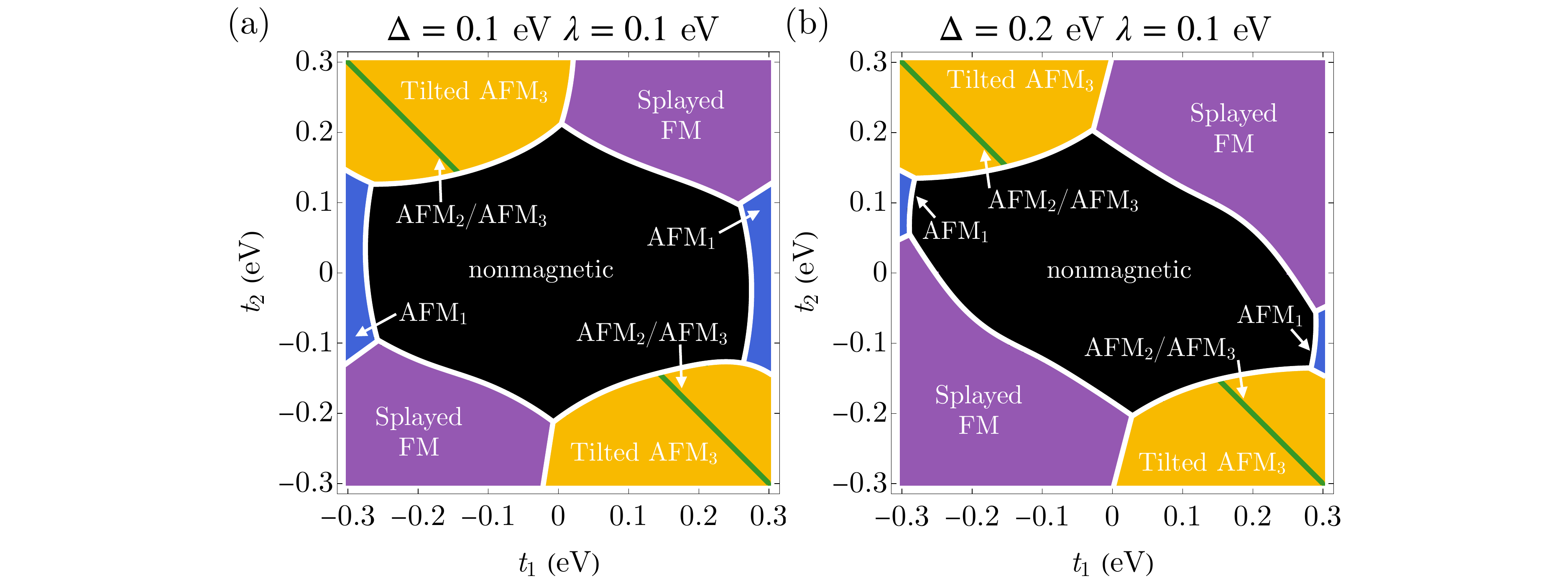}
    \caption{Phase diagram of the microscopic model with trigonal distortion: (a) The trigonal field $\Delta$ is comparable to spin--orbit coupling, (b) The trigonal field $\Delta$ is twice to spin--orbit coupling. All the other parameters are kept the same as in Fig.~\ref{fig:phasediagt1t2}(a).}
    \label{fig:A2Ru2O7}
\end{figure}

In the parameter regime dominated by oxygen-mediated hopping ($t_3=0$ and $t_4=0$), we find that a trigonal distortion of strength comparable to, or even larger than, the spin--orbit coupling does not qualitatively change the phase diagram from Fig.~\ref{fig:phasediagt1t2} (a) (see Fig.~\ref{fig:A2Ru2O7}). Using the DFT parameters from
Ref.~\cite{pyrochloreDFT2024} (listed in Table~\ref{tab:DFTparameters}), we find that  the A${}_2$Ru${}_2$O${}_7$ compounds lie predominantly in the upper-left region of the phase diagram in Fig.~\ref{fig:A2Ru2O7}, corresponding to experimentally measured AFM${}_2$/AFM${}_3$ or tilted AFM${}_3$ states (listed in Table~\ref{tab:A2Ru2O7}). While this comparison correctly captures the overall magnetic tendencies across the family, quantitative discrepancies remain. The calculated critical spin--orbit couplings (see Table~\ref{tab:DFTparameters}), above which the triplon condensate can no longer be stabilized and the $J=0$ singlet remains the ground state,
are generally smaller than the expected atomic value for Ru ions. Also the predicted magnetic ground states (last column of Table~\ref{tab:DFTparameters})
disagree with experiment for several compounds.
This suggests that the DFT-derived microscopic parameters require further refinement for a fully quantitative description.

\subsection*{Nd${}_2$Ru${}_2$O${}_7$ and Raman responses}

We now turn to Nd$_2$Ru$_2$O$_7$, for which inelastic Raman scattering measurements have revealed low-energy magnetic excitations~\cite{wulferding2023}. This compound provides an opportunity to test the microscopic Van Vleck framework against both the magnetic ground state and the excitation spectrum, including its Raman signatures.

 Starting from the hopping parameters obtained from DFT (Table~\ref{tab:DFTparameters} Ref.~\cite{pyrochloreDFT2024}), we find that the parameter set for Nd$_2$Ru$_2$O$_7$  lies very close to the excitonic quantum critical point, but remains slightly on the nonmagnetic side of the instability. Equivalently, the critical spin--orbit coupling required to stabilize the triplon-condensed phase is somewhat smaller than the realistic atomic value for Ru ion, $\lambda\simeq0.1$~eV. Since the spin-orbit coupling is an essentially 
atomic property of the Ru ion and is expected to vary only weakly among insulating ruthenates,
we attribute this discrepancy to the microscopic parameters inferred from DFT, whose values depend sensitively on structural details and electronic correlations. We therefore keep the physically realistic value $\lambda=0.1$~eV fixed and instead make only small adjustments to the hopping and interaction parameters.
Specifically, we increase $t_4$ from $0.034$ eV to $0.08$ eV while keeping all other parameters unchanged from the Nd$_2$Ru$_2$O$_7$ values listed in Table~\ref{tab:DFTparameters}. 
As shown in Table~\ref{tbl:hop}, the antisymmetric hopping $t_4$ originates from the absence of inversion symmetry on the Ru--Ru bond, making a moderate increase in $t_4$
reasonable in the pyrochlore lattice.

\begin{figure}
    \centering
    \includegraphics[width=0.7\linewidth]{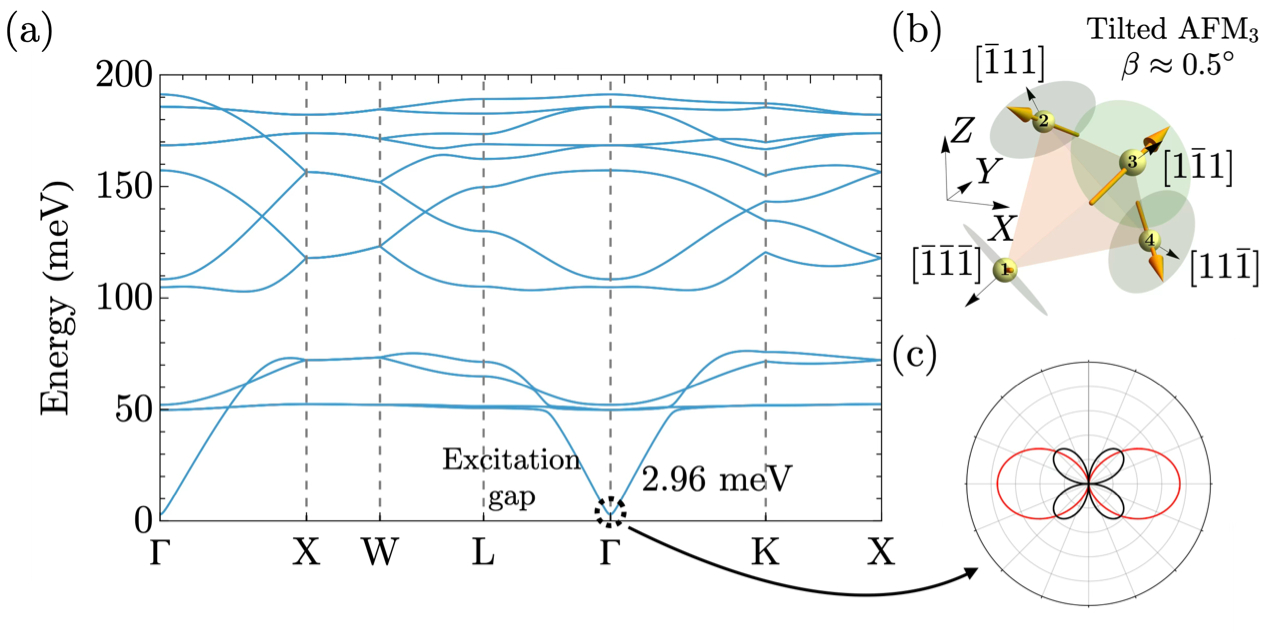}
    \caption{(a) Triplon excitation spectrum of Nd${}_2$Ru${}_2$O${}_7$ computed using the modified set of DFT parameters ($U_2=1.664$~eV, $J_H=0.344$~eV, $t_1=-0.1059$~eV, $t_2=0.1501$~eV, $t_3=0.00157$~eV, $t_4=0.08$~eV, $\lambda=0.1$~eV, and $\Delta=0.131$~eV) along with spin--orbit coupling $\lambda=0.1$~eV, chosen to stabilize the triplon condensate. (b) The resulting magnetic order is a tilted AFM${}_3$ state with a tilting angle $\beta\approx 0.5^\circ$, in which th Ru moments lie nearly within the degenerate plane of AFM${}_2$/AFM${}_3$ state. (c) Angle-resolved Raman reponse of the low-energy pseudo-Goldstone mode at the $\Gamma$ point, shown in the parallel (red) and perpendicular (black) polarization channels.}
    \label{fig:Nd2Ru2O7}
\end{figure}

The resulting parameter set yields a tilted AFM$_3$ ground state with magnetic moment $1.49~\mu_B$ and a small tilting angle $\beta\approx0.5^\circ$, in which the Ru moments lie nearly within the degenerate AFM$_2$/AFM$_3$ manifold [Fig.~\ref{fig:Nd2Ru2O7}(b)]. The corresponding low-energy excitation spectrum is shown in Fig.~\ref{fig:Nd2Ru2O7}(a), and the lowest magnetic excitation occurs at approximately $2.96$~meV, 
in agreement with the lowest-energy magnetic mode observed in Raman scattering~\cite{wulferding2023}, whose microscopic origin has remained unresolved.

To establish that this excitation indeed corresponds to the experimentally observed Raman mode, we compute its Raman intensity and polarization dependence.  Our aim here is not to describe the full Raman response of Nd$_2$Ru$_2$O$_7$, which has already been analyzed in Ref.~\cite{wulferding2023}, but rather to identify the microscopic origin of this previously unexplained low-energy magnetic mode. Although the
higher-energy excitations may hybridize with the $J=2$ manifold (neglected in the singlet--triplet picture of Van Vleck magnetism) when hopping becomes sufficiently strong, the ground state and lowest-energy excitation are energetically well separated from the $J=2$ manifold and should remain largely unaffected.
To this end, we evaluate the angle-resolved Raman intensity of the lowest excitation within the Loudon--Fleury formalism~\cite{LF1968} in the $xy$ plane for both parallel and crossed polarization channels [Fig.~\ref{fig:Nd2Ru2O7}(c)] (see the Methods section for details). The calculated polarization dependence agrees well with experiment~\cite{wulferding2023}, confirming that the observed low-energy Raman feature originates from the pseudo-Goldstone mode of the tilted AFM$_3$ state. The only noticeable discrepancy is the broken fourfold symmetry observed in the crossed polarization channel. This is likely a limitation of the Loudon--Fleury approximation, which neglects the effects of oxygen-mediated hopping and the inversion-symmetry-breaking hopping $t_4$. As discussed earlier, these processes can generate antisymmetric components in the Raman tensor therefore breaking the fourfold symmetry in the crossed polarization channel~\cite{Yang2021}.

\begin{table}[t]
    \centering
    \begin{tabular}{|c|c|c|c|c|}
        \hline
         &$T_N$ &  $\angle\text{Ru-O-Ru}$ & $\mu_{\mathrm{Ru}}$ & Ru magnetic structure \\\hline
         Y${}_2$Ru${}_2$O${}_7$ & $76(2)$~K~\cite{Kmiec2006} &$128.45(2)^\circ$~\cite{Kennedy1995} & $1.36\mu_B$ \cite{Ito2001} & AFM${}_{2}$/AFM${}_{3}$~\cite{Kmiec2006}\\\hline
         Pr${}_2$Ru${}_2$O${}_7$ & $163$~K~\cite{Zouari2009} &  $131.69^\circ$~\cite{Laurita2019} & $1.48\mu_B$~\cite{Duijn2017}& AFM${}_{2}$/AFM${}_{3}$~\cite{Duijn2017}\\\hline
         Nd${}_2$Ru${}_2$O${}_7$ & $146$~K~\cite{Ku2018} & $130.8^\circ$~\cite{Ku2018}& $1.18\mu_B$~\cite{Ito2001}&AFM${}_{2}$/AFM${}_{3}$~\cite{Ito2001}\\\hline
         Sm${}_2$Ru${}_2$O${}_7$ & $135$~K~\cite{Taira1999} & $128.85^\circ$~\cite{Pawar2017,Chakraborty2026} & / & /\\\hline
         Eu${}_2$Ru${}_2$O${}_7$ & $118$~K~\cite{Perez2015} &$125.52^\circ$~\cite{Chatterjee2024} & $2.79\mu_B$~\cite{Perez2013} & / \\\hline
         Gd${}_2$Ru${}_2$O${}_7$ &  $114(1)$~K\cite{Gurgul2007}& $130.23(1) ^\circ$~\cite{Castro2021} & $2.83\mu_B$~\cite{Gurgul2007}& tilted AFM${}_3$~\cite{Gurgul2007}  \\\hline
         Tb${}_2$Ru${}_2$O${}_7$ & $110$~K~\cite{Chang2010} &$128.48^\circ$~\cite{Kennedy1996} & $0.9(1)\mu_B$\cite{Chang2010}&tilted AFM${}_3$~\cite{Chang2010}\\\hline
         Dy${}_2$Ru${}_2$O${}_7$ &$\sim100$~K~\cite{Xu2014}& $130.23^\circ$~\cite{Yamamoto1994}& / &tilted AFM${}_3$~\cite{Xu2014}\\\hline
         Ho${}_2$Ru${}_2$O${}_7$ & $95$~K~\cite{Bansal2002} &$128.64^\circ$~\cite{Museur2026} & $1.2\mu_B$~\cite{Museur2026} & AFM${}_{2}$/AFM${}_{3}$~\cite{Museur2026}\\\hline
         Er${}_2$Ru${}_2$O${}_7$ & $95$~K~\cite{Taira2003}& $128.24(9)^\circ$~\cite{Taira2003}& $\sim2.0\mu_B$~\cite{Taira2003} & tilted AFM${}_3$~\cite{Taira2003}\\\hline
         Yb${}_2$Ru${}_2$O${}_7$ &$85$~K~\cite{Bustos2024}&  $126.9^\circ$~\cite{Bustos2024} &$1.41\mu_B$ \cite{Bustos2024} & AFM${}_{2}$/AFM${}_{3}$~\cite{Bustos2024} \\\hline
    \end{tabular}
    \caption{Experimentally reported properties of the Ru subsystem
    in the A${}_2$Ru${}_2$O${}_7$ family of compounds. The Ru-O-Ru bond angle is computed from the reported position of the O ion sitting at Wyckoff position $48f: (x,1/8,1/8)$ as $\arccos(1-2/(3+16x(2x-1)))$. A slash indicates that no experimental data have been reported in that entry.}
    \label{tab:A2Ru2O7}
\end{table}

\begin{table}[t]
    \centering
    \begin{tabular}{|c|c|c|c|c|c|c|c|c|c|}
        \hline
        & $U_2$ & $J_H$  & $t_1$ & $t_2$ & $t_3$ & $t_4$ & $\Delta$ & {\cbl $\lambda_c$} & {\cbl $|\Psi\rangle$} \\\hline
        Pr${}_2$Ru${}_2$O${}_7$ & $1.637$ & $0.341$ & $-0.1025$ & $0.1543$ & $0.0151$ & $0.0059$ &$0.130$ & \cbl{$0.1046$} & \cbl{tilted AFM${}_3$} \\\hline
        Nd${}_2$Ru${}_2$O${}_7$ & $1.664$ & $0.344$ & $-0.1059$ & $0.1501$ & $0.0157$ & $0.0034$ & $0.131$ & \cbl{$0.0977$} & \cbl{tilted AFM${}_3$}\\\hline
        Sm${}_2$Ru${}_2$O${}_7$ & $1.710$ & $0.349$ & $-0.1117$ & $0.1418$ & $0.0165$ & $-0.0012$ & $0.134$ & \cbl{$0.0848$} & \cbl{tilted AFM${}_3$}\\\hline
        Eu${}_2$Ru${}_2$O${}_7$ & $1.728$ & $0.351$ & $-0.1140$ & $0.1376$ & $0.0167$ & $-0.0032$ & $0.136$ & \cbl{$0.0786$} & \cbl{tilted AFM${}_3$}\\\hline
        Tb${}_2$Ru${}_2$O${}_7$ & $1.759$ & $0.353$ & $-0.1177$ & $0.1296$ & $0.0170$ & $-0.0068$ & $0.142$ & \cbl{$0.0672$} & \cbl{tilted AFM${}_3$}\\\hline
        Dy${}_2$Ru${}_2$O${}_7$ & $1.772$ & $0.355$ & $-0.1195$ & $0.1255$ & $0.0170$ & $-0.0085$ & $0.145$ & \cbl{$0.0616$} & \cbl{AIAO}\\\hline
        Ho${}_2$Ru${}_2$O${}_7$ & $1.782$ & $0.355$ & $-0.1211$ & $0.1212$ & $0.0171$ & $-0.0102$ & $0.149$ & \cbl{$0.0618$} & \cbl{AIAO}\\\hline
        Er${}_2$Ru${}_2$O${}_7$ & $1.793$ & $0.356$ & $-0.1225$ & $0.1170$ & $0.0171$ & $-0.0119$ & $0.152$ & \cbl{$0.0642$} & \cbl{AIAO}\\\hline
        Tm${}_2$Ru${}_2$O${}_7$ & $1.803$ & $0.357$ & $-0.1240$ & $0.1127$ & $0.0172$ & $-0.0135$ & $0.156$ & \cbl{$0.0667$} & \cbl{AIAO}\\\hline
        Yb${}_2$Ru${}_2$O${}_7$ & $1.811$ & $0.357$ & $-0.1247$ & $0.1091$ & $0.0170$ & $-0.0145$ & $0.158$ & \cbl{$0.0680$} & \cbl{AIAO}\\\hline
    \end{tabular}
    \caption{DFT parameters (in eV) for the A${}_2$Ru${}_2$O${}_7$ family of compounds~\cite{pyrochloreDFT2024}. The trigonal field $\Delta$ is extracted from Fig.~3(d) of Ref.~\cite{pyrochloreDFT2024}. From these parameters, we compute the critical spin--orbit coupling $\lambda_c$ (shown in blue) above which $J=0$ singlet is energetically favored as the ground state. The resulting condensate order (also in blue) is listed in the last column. }
    \label{tab:DFTparameters}
\end{table}

\section*{Discussion}

 In this work, we developed a microscopic theory of Van Vleck excitonic magnetism for pyrochlore ruthenates by deriving the effective singlet--triplet Hamiltonian directly from a multi-orbital Hubbard model with spin--orbit coupling. The resulting framework naturally explains how long-range magnetic order emerges from the condensation of spin--orbit excitons despite the local $J=0$ singlet ground state.

 The microscopic formulation also establishes a direct connection between the underlying electronic structure and the magnetic phase diagram. We identify the key hopping processes that govern the competition between different magnetic phases. The dominant oxygen-mediated hopping amplitudes, $t_1$ and $t_2$, determine the overall topology of the phase diagram, whereas the symmetry-allowed hopping channels $t_3$ and $t_4$ modify the phase boundaries and lift accidental degeneracies.

The resulting phase diagram reproduces all established $\mathbf{q}=0$ magnetic orders of the conventional spin-1 pyrochlore model~\cite{GangChen2018}. At the same time, the singlet--triplet structure of the excitonic model stabilizes a distinct tilted AFM$_3$ phase, which has no analogue in the pure spin-1 model and occupies an extended region of the phase diagram. 
For $t_3=t_4=0$, this phase lifts the accidental AFM$_2$/AFM$_3$ degeneracy everywhere except along the special line $t_1=-t_2$. Once either $t_3$ or $t_4$ becomes finite, the tilted AFM$_3$ phase is stabilized also along this line. A hallmark of this phase is a low-energy pseudo-Goldstone mode, whereas the remaining ordered phases are characterized by substantially larger excitation gaps.

More importantly, the microscopic theory establishes a direct connection between the electronic structure and experimentally observable collective excitations. Applying the Van Vleck excitonic magnetism framework to Nd$_2$Ru$_2$O$_7$, we showed that modest refinements of the DFT-derived microscopic parameters are sufficient to reproduce both the magnetic ground state and the energy of the lowest magnetic excitation observed in Raman scattering. The calculated Raman polarization dependence identifies this excitation as the pseudo-Goldstone mode of the tilted AFM$_3$ phase, thereby providing a microscopic explanation for a low-energy Raman feature whose origin had remained unresolved.

To conclude, we have demonstrated that pyrochlore ruthenates provide a concrete realization of the Van Vleck excitonic magnetism proposed by Khaliullin~\cite{khaliullin2013}. More broadly, our work establishes a microscopic route from the electronic structure of spin--orbit-coupled materials to their magnetic phases and spectroscopic signatures. Because it is formulated directly in terms of the underlying multi-orbital Hamiltonian, the approach developed here can be readily applied to other candidate excitonic magnets and extended to systems in which additional low-energy multiplets become relevant.


\section*{Methods}\label{sec:methods}

\subsection*{Local frames and coordinate transformation}\label{subsec:geometry}

The complexity of studying Van Vleck magnetism inside pyrochlore compounds lies in the locally rotated octahedral environment surrounding each Ru sublattice. The local orbital basis at each Ru site is defined by the orientation of the surrounding oxygen octahedron. In the ideal structure of pyrochlore, we define these axes in terms of the global coordinate:

\begin{equation}\label{localVecsEqn}
\begin{aligned}
\hat{x}_1&=\left(\frac{2}{3},-\frac{1}{3},\frac{2}{3}\right),&
\hat{y}_1&=\left(-\frac{1}{3},\frac{2}{3},\frac{2}{3}\right),&
\hat{z}_1&=\left(-\frac{2}{3},-\frac{2}{3},\frac{1}{3}\right),\\
\hat{x}_2&=\left(\frac{2}{3},\frac{1}{3},-\frac{2}{3}\right),&
\hat{y}_2&=\left(\frac{1}{3},\frac{2}{3},\frac{2}{3}\right),&
\hat{z}_2&=\left(\frac{2}{3},-\frac{2}{3},\frac{1}{3}\right),\\
\hat{x}_3&=\left(\frac{2}{3},\frac{1}{3},\frac{2}{3}\right),&
\hat{y}_3&=\left(\frac{1}{3},\frac{2}{3},-\frac{2}{3}\right),&
\hat{z}_3&=\left(-\frac{2}{3},\frac{2}{3},\frac{1}{3}\right),\\
\hat{x}_4&=\left(\frac{2}{3},-\frac{1}{3},-\frac{2}{3}\right),&
\hat{y}_4&=\left(-\frac{1}{3},\frac{2}{3},-\frac{2}{3}\right),&
\hat{z}_4&=\left(\frac{2}{3},\frac{2}{3},\frac{1}{3}\right).
\end{aligned}
\end{equation}

The $J=0$ singlet is invariant under rotation. The rotation of the $J=1$ triplet is implemented by the Wigner matrix
\begin{equation}
\mathcal{D}^{j}_{m'm}(\alpha,\beta,\gamma)
=
\braket{jm'|e^{-i\alpha \mathcal{J}_z}e^{-i\beta \mathcal{J}_y}e^{-i\gamma \mathcal{J}_z}|jm},
\end{equation}
where $\mathcal{J}_x$, $\mathcal{J}_y$, and $\mathcal{J}_z$ are spin-1 matrices. The Wigner matrix acts naturally in the angular-momentum basis 
$\{\ket{T_1},\ket{T_0},\ket{T_{\bar 1}}\}$ of the $J=1$ triplet. 
We denote the corresponding rotations from the local triplet basis to the global triplet basis at sublattice $\mu$ as $\mathcal{D}^{j=1}_{\mu}$. 
In terms of Euler angles,
\begin{equation}
\begin{aligned}
    (\alpha_1,\beta_1,\gamma_1)
=
\left(-\frac{3\pi}{4},\arctan(2\sqrt{2}),\frac{3\pi}{4}\right),
\qquad
(\alpha_2,\beta_2,\gamma_2)
=
\left(-\frac{\pi}{4},\arctan(2\sqrt{2}),\frac{\pi}{4}\right),\\
(\alpha_3,\beta_3,\gamma_3)
=
\left(\frac{3\pi}{4},\arctan(2\sqrt{2}),-\frac{3\pi}{4}\right),
\qquad
(\alpha_4,\beta_4,\gamma_4)
=
\left(\frac{\pi}{4},\arctan(2\sqrt{2}),-\frac{\pi}{4}\right).
\end{aligned}
\end{equation}
these rotations are described by the following Wigner matrices:
\begin{equation}
    \begin{aligned}
        \mathcal{D}_1^{j=1}&=\left(
        \begin{array}{ccc}
         \frac{2}{3} & \left(-\frac{1}{3}-\frac{i}{3}\right) \sqrt{2} & \frac{i}{3} \\
         \left(\frac{1}{3}-\frac{i}{3}\right) \sqrt{2} & \frac{1}{3} & \left(-\frac{1}{3}-\frac{i}{3}\right) \sqrt{2} \\
         -\frac{i}{3} & \left(\frac{1}{3}-\frac{i}{3}\right) \sqrt{2} & \frac{2}{3} \\
        \end{array}
        \right),\\
        \mathcal{D}_2^{j=1}&=\left(
        \begin{array}{ccc}
         \frac{2}{3} & \left(\frac{1}{3}-\frac{i}{3}\right) \sqrt{2} & -\frac{i}{3} \\
         \left(-\frac{1}{3}-\frac{i}{3}\right) \sqrt{2} & \frac{1}{3} & \left(\frac{1}{3}-\frac{i}{3}\right) \sqrt{2} \\
         \frac{i}{3} & \left(-\frac{1}{3}-\frac{i}{3}\right) \sqrt{2} & \frac{2}{3} \\
        \end{array}
        \right),\\
        \mathcal{D}_3^{j=1}&=\left(
        \begin{array}{ccc}
         \frac{2}{3} & \left(-\frac{1}{3}+\frac{i}{3}\right) \sqrt{2} & -\frac{i}{3} \\
         \left(\frac{1}{3}+\frac{i}{3}\right) \sqrt{2} & \frac{1}{3} & \left(-\frac{1}{3}+\frac{i}{3}\right) \sqrt{2} \\
         \frac{i}{3} & \left(\frac{1}{3}+\frac{i}{3}\right) \sqrt{2} & \frac{2}{3} \\
        \end{array}
        \right),\\
        \mathcal{D}_4^{j=1}&=\left(
        \begin{array}{ccc}
         \frac{2}{3} & \left(\frac{1}{3}+\frac{i}{3}\right) \sqrt{2} & \frac{i}{3} \\
         \left(-\frac{1}{3}+\frac{i}{3}\right) \sqrt{2} & \frac{1}{3} & \left(\frac{1}{3}+\frac{i}{3}\right) \sqrt{2} \\
         -\frac{i}{3} & \left(-\frac{1}{3}+\frac{i}{3}\right) \sqrt{2} & \frac{2}{3} \\
        \end{array}
        \right).
    \end{aligned}
\end{equation}

\subsection*{Construction of hopping matrices}
\label{subsec:hopping}

The hopping amplitudes appearing in Eq.~\eqref{eq:hopping} are defined in the local $t_{2g}$ orbital basis on each Ru site. To construct these hopping amplitudes,  we first define a global five-orbital basis for each sublattice $\mu$,
\begin{equation}
\ket{d_\mu^G}
=
\left(
d^G_{\mu,YZ},
d^G_{\mu,XZ},
d^G_{\mu,XY},
d^G_{\mu,X^2-Y^2},
d^G_{\mu,3Z^2-R^2}
\right)^T ,
\end{equation}
where the superscript $G$ indicates that the orbital shapes are expressed with respect to the global cubic axes.
The corresponding local five-orbital basis at sublattice $\mu$ is denoted by
\begin{equation}
\ket{d_\mu}
=
\left(
d_{\mu,yz},
d_{\mu,xz},
d_{\mu,xy},
d_{\mu,x^2-y^2},
d_{\mu,3z^2-r^2}
\right)^T ,
\end{equation}
where the orbital labels are defined with respect to the local axes 
$(\hat{x}_\mu,\hat{y}_\mu,\hat{z}_\mu)$. 
The transformation from the global to the local orbital basis is written as
\begin{equation}
\ket{d_\mu}=\mathcal{R}^\mu_{LG}\ket{d_\mu^G}.
\end{equation}

For a nearest-neighbor bond connected by sublattice $\mu$ and $\nu$ with direction cosines $(l,m,n)$, we denote by 
$\mathcal{T}^{G}_{\mu\nu}(l,m,n)$ the Slater--Koster hopping matrix \cite{Slater1954} written in the global five-orbital basis. 
Projecting this matrix to the local $t_{2g}$ bases at sublattices $\mu$ and $\nu$, respectively, gives
\begin{equation}
\mathcal{T}_{\mu\nu}^{t_{2g}}
=
\left[
\mathcal{R}^\mu_{LG}\,
\mathcal{T}^{G}_{ij}(l,m,n)\,
\left(\mathcal{R}^\nu_{LG}\right)^\dagger
\right]_{t_{2g}} .
\label{eq:t2g_hopping_construction}
\end{equation}

Because the local spin quantization axes differ from site to site, the spin part of the hopping must also be expressed in a common convention. These spin rotations $S_{\mu\nu}$ are computed by Wigner rotation matrices $\mathcal{D}_{\mu}^{j=1/2}(\alpha,\beta,\gamma)$ for the spin-$1/2$ degree of freedom:
\begin{align}
    S_{\mu\nu}=\mathcal{D}_{\nu}^{j=1/2\dagger}(\alpha_\nu,\beta_\nu,\gamma_\nu)\mathcal{D}_{\mu}^{j=1/2}(\alpha_\mu,\beta_\mu,\gamma_\mu).
\end{align}

Combining the orbital and spin parts, the hopping amplitudes entering Eq.~\eqref{eq:hopping} are
\begin{equation}
t_{ij}^{\alpha\sigma,\beta\sigma'}
=
\left(\mathcal{T}_{\mu\nu}^{t_{2g}}\right)_{\alpha\beta}
\left(S_{ij}\right)_{\sigma\sigma'},
\label{eq:full_hopping_matrix}
\end{equation}
where $\mu$,$\nu$ are sublattice indices of sites $i$ and $j$, respectively.
The explicit Slater--Koster expressions for $t_1,t_2,t_3,t_4$, together with the orbital rotations are given in Supplementary Sec.~\ref{slaterkoster}.

\subsection*{Projected superexchange Hamiltonian}
\label{subsec:projected_exchange}

The effective exchange Hamiltonian is obtained by second-order perturbation theory in the hopping amplitudes 
$t_{ij}^{\alpha\sigma,\beta\sigma'}$. 
For each nearest-neighbor bond, two successive hopping processes virtually take the system from the initial 
$d_i^4d_j^4$ manifold to intermediate charge-transfer configurations of the form $d_i^3d_j^5$, and then back to the $d_i^4d_j^4$ manifold. The single-site energy in this manifold is $E_4=6U_2-J_H$, so the initial two-site energy is $E_0=2E_4$.
The corresponding intermediate-state energy is $E_I=E_3+E_5$. 
Since the $d^5$ configuration has energy $10U_2$, while the $d^3$ configuration splits into three interaction channels, the relevant intermediate energies are
\begin{align}\label{3-5}
E_{I,1} &= 10U_2 + (3U_2 - 3J_H),\nonumber \\
E_{I,2} &= 10U_2 + 3U_2, \\
E_{I,3} &= 10U_2 + (3U_2 + J_H).\nonumber
\end{align}
After summing over the intermediate states, we project the resulting operator onto the local singlet-triplet manifold 
$\{\ket{s},\ket{T^x},\ket{T^y},\ket{T^z}\}$ on each site. In practice, we first compute the projected Hamiltonian for the representative $Z$ bond connecting sites 1 and 4. 
The corresponding two-site low-energy Hilbert space is spanned by
\begin{align}\label{eqn:twoSiteState}
\ket{\phi_{14,\alpha}}
\equiv
\ket{\tau_{1\mu}\rangle\otimes |\tau_{4\nu}},
\end{align}
where $\mu,\nu=0,1,2,3$ correspond to $|s\rangle,|T^x\rangle,|T^y\rangle,|T^z\rangle$ states, respectively. 
In this basis, the projected bond Hamiltonian is represented by a $16\times16$ matrix, denoted by $\mathcal{H}^Z_{14}$. 
The remaining nearest-neighbor bond Hamiltonians are generated from $H^Z_{14}$ by symmetry operations of the tetrahedral point group. 
These operations leave the singlet component invariant and permute the Cartesian triplet components. 
This procedure gives the six bond Hamiltonians on a tetrahedron,
\begin{equation}
\left\{
\mathcal{H}^Z_{14},\,
\mathcal{H}^X_{12},\,
\mathcal{H}^Y_{13},\,
\mathcal{H}^Z_{32},\,
\mathcal{H}^X_{43},\,
\mathcal{H}^Y_{42}
\right\}.
\end{equation}
The full nearest-neighbor exchange Hamiltonian is obtained by summing these bond Hamiltonians over all tetrahedra of the pyrochlore lattice, with the appropriate site labels. 
In the low-energy bosonic representation, this full interaction has the quartic form given in Eq.~\eqref{eq:Heff}, where the sum over $\langle ij\rangle$ runs over all nearest-neighbor bonds. 
The explicit permutation matrices and bond-generation relations are given in Supplementary Sec.~\ref{sec:symmetry_bonds}.

\subsection*{Magnetic Raman scattering}

Since Van Vleck magnetism originates from virtual hopping of electrons between Ru ions, we compute the magnetic Raman response using the Loudon--Fleury formalism~\cite{LF1968}. Within this approach, the Raman operator is generated by the same virtual hopping processes that give rise to the superexchange interaction. 
We define the Raman operator as
\begin{align}
    \mathcal{R}\equiv\sum_{\langle ij\rangle}\mathcal{R}_{\langle ij\rangle},
\end{align}
where $\mathcal{R}_{\langle ij\rangle}$ acts on a pair of neighboring Ru ions connected by the bond vector $\mathbf{d}_{ij}$ and takes the form
\begin{align}
    \mathcal{R}_{\langle ij\rangle}\equiv-(\boldsymbol{\epsilon}_{\mathrm{in}}\cdot\mathbf{d}_{ij})(\boldsymbol{\epsilon}_{\mathrm{out}}\cdot\mathbf{d}_{ij})\mathcal{H}_{\mathrm{eff}}^{(4)}.
\end{align}
Here $\mathcal{H}_{\rm eff}^{(4)}$ is the superexchange Hamiltonian [Eq.~\eqref{eq:Heff}] obtained from second-order perturbation theory in the virtual hopping, while $\boldsymbol{\epsilon}_{\rm in(out)}$ denote the polarization vectors of the incoming (outgoing) photons. We assume the incoming light frequency to be off-resonance, so that the Raman operator is approximately frequency independent.

The Raman intensity is given by the dynamical correlation function of the Raman operator,
\begin{align}
    I(\Omega)\propto \int dt\, e^{i\Omega t}
    \langle \mathcal{R}(t)\mathcal{R}(0)\rangle.
\end{align}
To compute the angle-resolved Raman response, we consider the parallel channel $\boldsymbol{\epsilon}_{\mathrm{in}}^\parallel=\boldsymbol{\epsilon}_{\mathrm{out}}^\parallel=(\cos\theta,\sin\theta,0)$ and the crossed channel $\boldsymbol{\epsilon}_{\mathrm{in}}^\perp=(\cos\theta,\sin\theta,0)$, $\boldsymbol{\epsilon}_{\mathrm{out}}^\perp=(-\sin\theta,\cos\theta,0)$, 
where the polarization vectors lie in the global $XY$-plane.
In the triplon-condensed phase, the elementary excitations are no longer the original triplons $\{|T^x\rangle,|T^y\rangle,|T^z\rangle\}$ but the normal modes $\{|\Phi_1\rangle,|\Phi_2\rangle,|\Phi_3\rangle\}$ obtained after the unitary transformation of Eq.~\eqref{eq:unitary}. Accordingly, the superexchange Hamiltonian, and hence the Raman operator, is rewritten in this basis and expanded to quadratic order in the excitation operators. The Raman spectra are then obtained from the corresponding quadratic Hamiltonian.


\section*{Acknowledgments}
The authors thank Giniyat Khaliullin and Ioannis Rousochatzakis for valuable discussions on the mechanism of Van Vleck excitonic magnetism. Y.Y. and N.B.P. also acknowledge Dirk Wulferding for a fruitful collaboration on the Raman scattering studies of Nd$_2$Ru$_2$O$_7$, which motivated part of this work. S.S., Y.Y, and N.B.P. acknowledge the support from NSF DMR-2310318 and the
support of the Minnesota Supercomputing Institute (MSI) at
the University of Minnesota.  N.B.P. also  acknowledges the hospitality of the Aspen Center for Physics, which is supported by National Science Foundation grant PHY-2210452.

\section*{Author Contributions}
S.S. and Y.Y. contributed equally to this work and share first authorship.
Y.Y. and N.B.P. devised the project.
S.S., Y.Y. and N.B.P. performed  calculations. Y.Y. performed the analysis of the Raman response in Nd$_2$Ru$_2$O$_7$.
 S.S., Y.Y. and N.B.P. contributed to the interpretation of the results and the writing of the paper.

\section*{Competing interests}
The authors declare no competing interests.

\section*{Data availability}
Most of the calculations presented in this work are analytical and are described in the Methods and Supplementary Information. The theoretical data generated and analyzed during this study are not publicly available because they primarily consist of intermediate analytical calculations and numerical checks that are not required to reproduce the main results. They are, however, available from the corresponding author.

\section*{Code availability}
The mathematica codes used during the current study are not publicly available because they contain preliminary and intermediate calculations that are not directly essential for reproducing or interpreting the main results, but are available from the corresponding author upon reasonable request.

\bibliography{refs}

\begin{thebibliography}{10}
\expandafter\ifx\csname url\endcsname\relax
  \def\url#1{\texttt{#1}}\fi
\expandafter\ifx\csname urlprefix\endcsname\relax\def\urlprefix{URL }\fi
\providecommand{\bibinfo}[2]{#2}
\providecommand{\eprint}[2][]{\url{#2}}

\bibitem{khaliullin2013}
\bibinfo{author}{Khaliullin, G.}
\newblock \bibinfo{title}{{Excitonic Magnetism in Van Vleck--type ${d}^{4}$
  Mott Insulators}}.
\newblock \emph{\bibinfo{journal}{Phys. Rev. Lett.}}
  \textbf{\bibinfo{volume}{111}}, \bibinfo{pages}{197201}
  (\bibinfo{year}{2013}).

\bibitem{Akbari2014}
\bibinfo{author}{Akbari, A.} \& \bibinfo{author}{Khaliullin, G.}
\newblock \bibinfo{title}{{Magnetic excitations in a spin-orbit-coupled
  ${d}^{4}$ Mott insulator on the square lattice}}.
\newblock \emph{\bibinfo{journal}{Phys. Rev. B}} \textbf{\bibinfo{volume}{90}},
  \bibinfo{pages}{035137} (\bibinfo{year}{2014}).

\bibitem{Souliou2017}
\bibinfo{author}{Souliou, S.-M.} \emph{et~al.}
\newblock \bibinfo{title}{{Raman Scattering from Higgs Mode Oscillations in the
  Two-Dimensional Antiferromagnet ${\mathrm{Ca}}_{2}{\mathrm{RuO}}_{4}$}}.
\newblock \emph{\bibinfo{journal}{Phys. Rev. Lett.}}
  \textbf{\bibinfo{volume}{119}}, \bibinfo{pages}{067201}
  (\bibinfo{year}{2017}).

\bibitem{Jain2017}
\bibinfo{author}{Jain, A.} \emph{et~al.}
\newblock \bibinfo{title}{{Higgs mode and its decay in a two-dimensional
  antiferromagnet}}.
\newblock \emph{\bibinfo{journal}{Nature Physics}}
  \textbf{\bibinfo{volume}{13}}, \bibinfo{pages}{633--637}
  (\bibinfo{year}{2017}).

\bibitem{Gretarsson2019}
\bibinfo{author}{Gretarsson, H.} \emph{et~al.}
\newblock \bibinfo{title}{{Observation of spin-orbit excitations and Hund's
  multiplets in ${\mathrm{Ca}}_{2}{\mathrm{RuO}}_{4}$}}.
\newblock \emph{\bibinfo{journal}{Phys. Rev. B}}
  \textbf{\bibinfo{volume}{100}}, \bibinfo{pages}{045123}
  (\bibinfo{year}{2019}).

\bibitem{chaloupka2019highly}
\bibinfo{author}{Chaloupka, J.} \& \bibinfo{author}{Khaliullin, G.}
\newblock \bibinfo{title}{Highly frustrated magnetism in relativistic d 4 mott
  insulators: Bosonic analog of the kitaev honeycomb model}.
\newblock \emph{\bibinfo{journal}{Physical Review B}}
  \textbf{\bibinfo{volume}{100}}, \bibinfo{pages}{224413}
  (\bibinfo{year}{2019}).

\bibitem{Takahashi2021}
\bibinfo{author}{Takahashi, H.} \emph{et~al.}
\newblock \bibinfo{title}{{Nonmagnetic $J=0$ State and Spin-Orbit Excitations
  in ${\mathrm{K}}_{2}{\mathrm{RuCl}}_{6}$}}.
\newblock \emph{\bibinfo{journal}{Phys. Rev. Lett.}}
  \textbf{\bibinfo{volume}{127}}, \bibinfo{pages}{227201}
  (\bibinfo{year}{2021}).

\bibitem{singlet2026}
\bibinfo{author}{Pätzold, L.} \emph{et~al.}
\newblock \bibinfo{title}{{Between Mott and cluster Mott: spin-orbit entangled
  dimer singlets in Ba$_3$CeRu$_2$O$_9$}}  (\bibinfo{year}{2026}).
\newblock \eprint{2604.06886}.

\bibitem{KhaliullinPRL2016}
\bibinfo{author}{Chaloupka, J. c.~v.} \& \bibinfo{author}{Khaliullin, G.}
\newblock \bibinfo{title}{{Doping-Induced Ferromagnetism and Possible Triplet
  Pairing in ${d}^{4}$ Mott Insulators}}.
\newblock \emph{\bibinfo{journal}{Phys. Rev. Lett.}}
  \textbf{\bibinfo{volume}{116}}, \bibinfo{pages}{017203}
  (\bibinfo{year}{2016}).

\bibitem{Cao2014}
\bibinfo{author}{Cao, G.} \emph{et~al.}
\newblock \bibinfo{title}{{Novel Magnetism of Ir$^{5+}$ ($5d^4$) Ions in the
  Double Perovskite Sr$_2$YIrO$_6$}}.
\newblock \emph{\bibinfo{journal}{Physical Review Letters}}
  \textbf{\bibinfo{volume}{112}}, \bibinfo{pages}{056402}
  (\bibinfo{year}{2014}).

\bibitem{LagunaPRB2020}
\bibinfo{author}{Laguna-Marco, M.~A.} \emph{et~al.}
\newblock \bibinfo{title}{{Magnetism of ${\mathrm{Ir}}^{5+}$-based double
  perovskites: Unraveling its nature and the influence of structure}}.
\newblock \emph{\bibinfo{journal}{Phys. Rev. B}}
  \textbf{\bibinfo{volume}{101}}, \bibinfo{pages}{014449}
  (\bibinfo{year}{2020}).

\bibitem{ChenPRB2017}
\bibinfo{author}{Chen, Q.} \emph{et~al.}
\newblock \bibinfo{title}{{Magnetism out of antisite disorder in the $J=0$
  compound ${\mathrm{Ba}}_{2}{\mathrm{YIrO}}_{6}$}}.
\newblock \emph{\bibinfo{journal}{Phys. Rev. B}} \textbf{\bibinfo{volume}{96}},
  \bibinfo{pages}{144423} (\bibinfo{year}{2017}).

\bibitem{DeyPRB2016}
\bibinfo{author}{Dey, T.} \emph{et~al.}
\newblock \bibinfo{title}{{${\text{Ba}}_{2}{\text{YIrO}}_{6}$: A cubic double
  perovskite material with ${\text{Ir}}^{5+}$ ions}}.
\newblock \emph{\bibinfo{journal}{Phys. Rev. B}} \textbf{\bibinfo{volume}{93}},
  \bibinfo{pages}{014434} (\bibinfo{year}{2016}).

\bibitem{Schnait2022}
\bibinfo{author}{Schnait, H.}, \bibinfo{author}{Bauernfeind, D.},
  \bibinfo{author}{Saha-Dasgupta, T.} \& \bibinfo{author}{Aichhorn, M.}
\newblock \bibinfo{title}{{Small moments without long-range magnetic ordering
  in the zero-temperature ground state of the double perovskite iridate
  ${\mathrm{Ba}}_{2}{\mathrm{YIrO}}_{6}$}}.
\newblock \emph{\bibinfo{journal}{Phys. Rev. B}}
  \textbf{\bibinfo{volume}{106}}, \bibinfo{pages}{035132}
  (\bibinfo{year}{2022}).

\bibitem{KuschPRB2018}
\bibinfo{author}{Kusch, M.} \emph{et~al.}
\newblock \bibinfo{title}{{Observation of heavy spin-orbit excitons propagating
  in a nonmagnetic background: The case of
  ${(\mathrm{Ba},\mathrm{Sr})}_{2}{\mathrm{YIrO}}_{6}$}}.
\newblock \emph{\bibinfo{journal}{Phys. Rev. B}} \textbf{\bibinfo{volume}{97}},
  \bibinfo{pages}{064421} (\bibinfo{year}{2018}).

\bibitem{PajskrPRB2016}
\bibinfo{author}{Pajskr, K.} \emph{et~al.}
\newblock \bibinfo{title}{{On the possibility of excitonic magnetism in Ir
  double perovskites}}.
\newblock \emph{\bibinfo{journal}{Phys. Rev. B}} \textbf{\bibinfo{volume}{93}},
  \bibinfo{pages}{035129} (\bibinfo{year}{2016}).

\bibitem{WilsonPRB2022}
\bibinfo{author}{Porter, Z.} \emph{et~al.}
\newblock \bibinfo{title}{{Spin-orbit excitons and electronic configuration of
  the $5{d}^{4}$ insulator
  ${\mathrm{Sr}}_{3}{\mathrm{Ir}}_{2}{\mathrm{O}}_{7}{\mathrm{F}}_{2}$}}.
\newblock \emph{\bibinfo{journal}{Phys. Rev. B}}
  \textbf{\bibinfo{volume}{106}}, \bibinfo{pages}{115140}
  (\bibinfo{year}{2022}).

\bibitem{AczelPRR2022}
\bibinfo{author}{Aczel, A.~A.} \emph{et~al.}
\newblock \bibinfo{title}{{Spin-orbit coupling controlled ground states in the
  double perovskite iridates ${A}_{2}B{\mathrm{IrO}}_{6}$ ($A=$ Ba, Sr; $B=$
  Lu, Sc)}}.
\newblock \emph{\bibinfo{journal}{Phys. Rev. Mater.}}
  \textbf{\bibinfo{volume}{6}}, \bibinfo{pages}{094409} (\bibinfo{year}{2022}).

\bibitem{Taira1999}
\bibinfo{author}{Taira, N.}, \bibinfo{author}{Wakeshima, M.} \&
  \bibinfo{author}{Hinatsu, Y.}
\newblock \bibinfo{title}{{Magnetic properties of ruthenium pyrochlores
  {R$_2$Ru$_2$O$_7$} ({R} = rare earth)}}.
\newblock \emph{\bibinfo{journal}{Journal of Physics: Condensed Matter}}
  \textbf{\bibinfo{volume}{11}}, \bibinfo{pages}{6983--6994}
  (\bibinfo{year}{1999}).

\bibitem{Ito2001}
\bibinfo{author}{Ito, M.} \emph{et~al.}
\newblock \bibinfo{title}{{Nature of spin freezing transition of geometrically
  frustrated pyrochlore system $R_2$Ru$_2$O$_7$ ($R$ = rare earth elements and
  Y)}}.
\newblock \emph{\bibinfo{journal}{Journal of Physics and Chemistry of Solids}}
  \textbf{\bibinfo{volume}{62}}, \bibinfo{pages}{337--341}
  (\bibinfo{year}{2001}).

\bibitem{Kennedy1995}
\bibinfo{author}{Kennedy, B.~J.}
\newblock \bibinfo{title}{{Structure Refinement of Y${}_2$Ru${}_2$O${}_7$ by
  Neutron Powder Diffraction}}.
\newblock \emph{\bibinfo{journal}{Acta Crystallographica Section C}}
  \textbf{\bibinfo{volume}{51}}, \bibinfo{pages}{790--792}
  (\bibinfo{year}{1995}).

\bibitem{Ku2018}
\bibinfo{author}{Ku, S.~T.} \emph{et~al.}
\newblock \bibinfo{title}{{Low temperature magnetic properties of
  Nd$_2$Ru$_2$O$_7$}}.
\newblock \emph{\bibinfo{journal}{Journal of Physics: Condensed Matter}}
  \textbf{\bibinfo{volume}{30}}, \bibinfo{pages}{155601}
  (\bibinfo{year}{2018}).

\bibitem{Laurita2019}
\bibinfo{author}{Laurita, G.} \emph{et~al.}
\newblock \bibinfo{title}{{Uncorrelated Bi off-centering and the
  insulator-to-metal transition in ruthenium
  ${A}_{2}{\mathrm{Ru}}_{2}{\mathrm{O}}_{7}$ pyrochlores}}.
\newblock \emph{\bibinfo{journal}{Phys. Rev. Mater.}}
  \textbf{\bibinfo{volume}{3}}, \bibinfo{pages}{095003} (\bibinfo{year}{2019}).

\bibitem{wulferding2023}
\bibinfo{author}{Wulferding, D.} \emph{et~al.}
\newblock \bibinfo{title}{{Collective magnetic Higgs excitation in a pyrochlore
  ruthenate}}.
\newblock \emph{\bibinfo{journal}{npj Quantum Materials}}
  \textbf{\bibinfo{volume}{8}}, \bibinfo{pages}{40} (\bibinfo{year}{2023}).

\bibitem{Lee2023}
\bibinfo{author}{Lee, J.~H.} \emph{et~al.}
\newblock \bibinfo{title}{{Linear scaling relationship of N\'eel temperature
  and dominant magnons in pyrochlore ruthenates}}.
\newblock \emph{\bibinfo{journal}{Phys. Rev. B}}
  \textbf{\bibinfo{volume}{108}}, \bibinfo{pages}{054443}
  (\bibinfo{year}{2023}).

\bibitem{pyrochloreDFT2024}
\bibinfo{author}{Li, J.} \emph{et~al.}
\newblock \bibinfo{title}{{Variation of electron-electron interaction in
  pyrochlore structures}}.
\newblock \emph{\bibinfo{journal}{Physical Review B}}
  \textbf{\bibinfo{volume}{110}}, \bibinfo{pages}{245147}
  (\bibinfo{year}{2024}).

\bibitem{Kmiec2006}
\bibinfo{author}{{Kmie\ifmmode \acute{c}\else \'{c}\fi{}, R. and \ifmmode
  \acute{S}\else \'{S}\fi{}wi\k{a}tkowska, \ifmmode \dot{Z}\else \.{Z}\fi{}.
  and Gurgul, J. and Rams, M. and Zarzycki, A. and Tomala, K.}}
\newblock \bibinfo{title}{{Investigation of the magnetic properties of
  ${\mathrm{Y}}_{2}{\mathrm{Ru}}_{2}{\mathrm{O}}_{7}$ by $^{99}\mathrm{Ru}$
  M\"ossbauer spectroscopy}}.
\newblock \emph{\bibinfo{journal}{Phys. Rev. B}} \textbf{\bibinfo{volume}{74}},
  \bibinfo{pages}{104425} (\bibinfo{year}{2006}).

\bibitem{Zouari2009}
\bibinfo{author}{Zouari, S.}, \bibinfo{author}{Ballou, R.},
  \bibinfo{author}{Cheikhrouhou, A.} \& \bibinfo{author}{Strobel, P.}
\newblock \bibinfo{title}{{Structural and magnetic properties of the
  (Bi${}_{2-x}$Pr${}_x$)Ru${}_2$O${}_7$ pyrochlore solid solution ($0\leq x\leq
  2$)}}.
\newblock \emph{\bibinfo{journal}{Journal of Alloys and Compounds}}
  \textbf{\bibinfo{volume}{476}}, \bibinfo{pages}{43--48}
  (\bibinfo{year}{2009}).

\bibitem{Perez2015}
\bibinfo{author}{Muñoz~Pérez, S.} \emph{et~al.}
\newblock \bibinfo{title}{Ruthenium-europium configuration in the
  eu${}_2$ru${}_2$o${}_7$ pyrochlore}.
\newblock \emph{\bibinfo{journal}{Journal of Applied Physics}}
  \textbf{\bibinfo{volume}{117}}, \bibinfo{pages}{17C702}
  (\bibinfo{year}{2015}).
\newblock
  \eprint{https://pubs.aip.org/aip/jap/article-pdf/doi/10.1063/1.4906528/15160921/17c702_1_online.pdf}.

\bibitem{Gurgul2007}
\bibinfo{author}{{Gurgul, J. and Rams, M. and \ifmmode \acute{S}\else
  \'{S}\fi{}wi\k{a}tkowska, \ifmmode \dot{Z}\else \.{Z}\fi{}. and Kmie\ifmmode
  \acute{c}\else \'{c}\fi{}, R. and Tomala, K.}}
\newblock \bibinfo{title}{{Bulk magnetic measurements and $^{99}\mathrm{Ru}$
  and $^{155}\mathrm{Gd}$ M\"ossbauer spectroscopies of
  ${\mathrm{Gd}}_{2}{\mathrm{Ru}}_{2}{\mathrm{O}}_{7}$}}.
\newblock \emph{\bibinfo{journal}{Phys. Rev. B}} \textbf{\bibinfo{volume}{75}},
  \bibinfo{pages}{064426} (\bibinfo{year}{2007}).

\bibitem{Chang2010}
\bibinfo{author}{Chang, L.~J.} \emph{et~al.}
\newblock \bibinfo{title}{{Magnetic order in the double pyrochlore
  Tb${}_2$Ru${}_2$O${}_7$}}.
\newblock \emph{\bibinfo{journal}{Journal of Physics: Condensed Matter}}
  \textbf{\bibinfo{volume}{22}}, \bibinfo{pages}{076003}
  (\bibinfo{year}{2010}).

\bibitem{Bustos2024}
\bibinfo{author}{{Ruiz Bustos}, R.}, \bibinfo{author}{{van Duijn}, J.},
  \bibinfo{author}{Lamura, G.}, \bibinfo{author}{Manuel, P.} \&
  \bibinfo{author}{Sanna, S.}
\newblock \bibinfo{title}{{Magnetic ordering in the frustrated pyrochlore
  Yb${}_2$Ru${}_2$O${}_7$}}.
\newblock \emph{\bibinfo{journal}{Journal of Alloys and Compounds}}
  \textbf{\bibinfo{volume}{1008}}, \bibinfo{pages}{176661}
  (\bibinfo{year}{2024}).

\bibitem{GangChen2018}
\bibinfo{author}{Li, F.-Y.} \& \bibinfo{author}{Chen, G.}
\newblock \bibinfo{title}{Competing phases and topological excitations of
  spin-1 pyrochlore antiferromagnets}.
\newblock \emph{\bibinfo{journal}{Phys. Rev. B}} \textbf{\bibinfo{volume}{98}},
  \bibinfo{pages}{045109} (\bibinfo{year}{2018}).

\bibitem{AbragamBleaney1970}
\bibinfo{author}{Abragam, A.} \& \bibinfo{author}{Bleaney, B.}
\newblock \emph{\bibinfo{title}{{Electron Paramagnetic Resonance of Transition
  Ions}}} (\bibinfo{publisher}{Clarendon Press}, \bibinfo{address}{Oxford},
  \bibinfo{year}{1970}).

\bibitem{remund2022semi}
\bibinfo{author}{Remund, K.}, \bibinfo{author}{Pohle, R.},
  \bibinfo{author}{Akagi, Y.}, \bibinfo{author}{Romh{\'a}nyi, J.} \&
  \bibinfo{author}{Shannon, N.}
\newblock \bibinfo{title}{{Semi-classical simulation of spin-1 magnets}}.
\newblock \emph{\bibinfo{journal}{Physical Review Research}}
  \textbf{\bibinfo{volume}{4}}, \bibinfo{pages}{033106} (\bibinfo{year}{2022}).

\bibitem{Pawar2017}
\bibinfo{author}{Pawar, R.} \emph{et~al.}
\newblock \bibinfo{title}{{Chemical synthesis and characterization of
  nano-sized rare-earth ruthenium pyrochlore compounds Ln${}_2$Ru${}_2$O${}_7$
  (Ln= rare earth)}}.
\newblock \emph{\bibinfo{journal}{Bulletin of Materials Science}}
  \textbf{\bibinfo{volume}{40}}, \bibinfo{pages}{1335--1345}
  (\bibinfo{year}{2017}).

\bibitem{Chakraborty2026}
\bibinfo{author}{Chakraborty, D.}, \bibinfo{author}{Maruthamuthu, S.},
  \bibinfo{author}{P, S.~K.}, \bibinfo{author}{Saravanakumar, B.} \&
  \bibinfo{author}{Vijayakumar, E.}
\newblock \bibinfo{title}{Tailored electrochemical properties by defect
  engineered pyrochlore structured samarium ruthenate}.
\newblock \emph{\bibinfo{journal}{Applied Physics A}}
  \textbf{\bibinfo{volume}{132}}, \bibinfo{pages}{537} (\bibinfo{year}{2026}).

\bibitem{Chatterjee2024}
\bibinfo{author}{Chatterjee, S.} \& \bibinfo{author}{Das, I.}
\newblock \bibinfo{title}{Desertion of anomalous magnetic transition and
  emergence of metallic state in cu doped eu${}_2$ru${}_2$o${}_7$ pyrochlore}.
\newblock \emph{\bibinfo{journal}{The Journal of Chemical Physics}}
  \textbf{\bibinfo{volume}{161}}, \bibinfo{pages}{244706}
  (\bibinfo{year}{2024}).

\bibitem{Castro2021}
\bibinfo{author}{Castro, A.}, \bibinfo{author}{Rosas-Huerta, J.} \&
  \bibinfo{author}{Escamilla, R.}
\newblock \bibinfo{title}{{Effect of Mo substitution on the structure and
  electrical properties of Gd${}_2$Ru${}_2$O${}_7$ pyrochlore}}.
\newblock \emph{\bibinfo{journal}{Physica B: Condensed Matter}}
  \textbf{\bibinfo{volume}{619}}, \bibinfo{pages}{413227}
  (\bibinfo{year}{2021}).

\bibitem{Kennedy1996}
\bibinfo{author}{Kennedy, B.} \& \bibinfo{author}{Vogt, T.}
\newblock \bibinfo{title}{Structural and bonding trends in ruthenium
  pyrochlores}.
\newblock \emph{\bibinfo{journal}{Journal of Solid State Chemistry}}
  \textbf{\bibinfo{volume}{126}}, \bibinfo{pages}{261--270}
  (\bibinfo{year}{1996}).

\bibitem{Yamamoto1994}
\bibinfo{author}{Yamamoto, T.} \emph{et~al.}
\newblock \bibinfo{title}{{Crystal Structure and Metal-Semiconductor Transition
  of the Bi${}_{2-x}$Ln${}_x$Ru${}_2$O${}_7$ Pyrochlores (Ln = Pr-Lu)}}.
\newblock \emph{\bibinfo{journal}{Journal of Solid State Chemistry}}
  \textbf{\bibinfo{volume}{109}}, \bibinfo{pages}{372--383}
  (\bibinfo{year}{1994}).

\bibitem{Museur2026}
\bibinfo{author}{Museur, F.} \emph{et~al.}
\newblock \bibinfo{title}{{Ferromagnetic fragmented state in the pyrochlore
  ${\mathrm{Ho}}_{2}{\mathrm{Ru}}_{2}{\mathrm{O}}_{7}$}}.
\newblock \emph{\bibinfo{journal}{Phys. Rev. B}}
  \textbf{\bibinfo{volume}{113}}, \bibinfo{pages}{L060406}
  (\bibinfo{year}{2026}).

\bibitem{Taira2003}
\bibinfo{author}{Taira, N.}, \bibinfo{author}{Wakeshima, M.},
  \bibinfo{author}{Hinatsu, Y.}, \bibinfo{author}{Tobo, A.} \&
  \bibinfo{author}{Ohoyama, K.}
\newblock \bibinfo{title}{{Magnetic structure of pyrochlore-type
  Er${}_2$Ru${}_2$O${}_7$}}.
\newblock \emph{\bibinfo{journal}{Journal of Solid State Chemistry}}
  \textbf{\bibinfo{volume}{176}}, \bibinfo{pages}{165--169}
  (\bibinfo{year}{2003}).

\bibitem{Kugel1982}
\bibinfo{author}{Kugel', K.~I.} \& \bibinfo{author}{Khomskiĭ, D.~I.}
\newblock \bibinfo{title}{The jahn-teller effect and magnetism: transition
  metal compounds}.
\newblock \emph{\bibinfo{journal}{Soviet Physics Uspekhi}}
  \textbf{\bibinfo{volume}{25}}, \bibinfo{pages}{231} (\bibinfo{year}{1982}).

\bibitem{Huebsch2022}
\bibinfo{author}{Huebsch, M.-T.}, \bibinfo{author}{Nomura, Y.},
  \bibinfo{author}{Sakai, S.} \& \bibinfo{author}{Arita, R.}
\newblock \bibinfo{title}{Magnetic structures and electronic properties of
  cubic-pyrochlore ruthenates from first principles}.
\newblock \emph{\bibinfo{journal}{Journal of Physics: Condensed Matter}}
  \textbf{\bibinfo{volume}{34}}, \bibinfo{pages}{194003}
  (\bibinfo{year}{2022}).

\bibitem{LF1968}
\bibinfo{author}{Fleury, P.~A.} \& \bibinfo{author}{Loudon, R.}
\newblock \bibinfo{title}{{Scattering of Light by One- and Two-Magnon
  Excitations}}.
\newblock \emph{\bibinfo{journal}{Phys. Rev.}} \textbf{\bibinfo{volume}{166}},
  \bibinfo{pages}{514--530} (\bibinfo{year}{1968}).

\bibitem{Yang2021}
\bibinfo{author}{Yang, Y.}, \bibinfo{author}{Li, M.},
  \bibinfo{author}{Rousochatzakis, I.} \& \bibinfo{author}{Perkins, N.~B.}
\newblock \bibinfo{title}{Non-loudon-fleury raman scattering in spin-orbit
  coupled mott insulators}.
\newblock \emph{\bibinfo{journal}{Phys. Rev. B}}
  \textbf{\bibinfo{volume}{104}}, \bibinfo{pages}{144412}
  (\bibinfo{year}{2021}).

\bibitem{Duijn2017}
\bibinfo{author}{Van~Duijn, J.} \emph{et~al.}
\newblock \bibinfo{title}{Induced quadrupolar singlet ground state of
  praseodymium in a modulated pyrochlore}.
\newblock \emph{\bibinfo{journal}{Phys. Rev. B}} \textbf{\bibinfo{volume}{96}},
  \bibinfo{pages}{094409} (\bibinfo{year}{2017}).

\bibitem{Perez2013}
\bibinfo{author}{Muñoz~Pérez, S.} \emph{et~al.}
\newblock \bibinfo{title}{{Anomalous electronic and magnetic properties of the
  Eu${}_2$Ru${}_2$O${}_7$ pyrochlore}}.
\newblock \emph{\bibinfo{journal}{Journal of Applied Physics}}
  \textbf{\bibinfo{volume}{113}}, \bibinfo{pages}{17E102}
  (\bibinfo{year}{2013}).

\bibitem{Xu2014}
\bibinfo{author}{Xu, Z.-C.} \emph{et~al.}
\newblock \bibinfo{title}{{Experimental observations of ferroelectricity in
  double pyrochlore Dy${}_2$Ru${}_2$O${}_7$}}.
\newblock \emph{\bibinfo{journal}{Frontiers of Physics}}
  \textbf{\bibinfo{volume}{9}}, \bibinfo{pages}{82--89} (\bibinfo{year}{2014}).

\bibitem{Bansal2002}
\bibinfo{author}{Bansal, C.}, \bibinfo{author}{Kawanaka, H.},
  \bibinfo{author}{Bando, H.} \& \bibinfo{author}{Nishihara, Y.}
\newblock \bibinfo{title}{{Structure and magnetic properties of the pyrochlore
  ${\mathrm{Ho}}_{2}{\mathrm{Ru}}_{2}{\mathrm{O}}_{7}:$ A possible dipolar spin
  ice system}}.
\newblock \emph{\bibinfo{journal}{Phys. Rev. B}} \textbf{\bibinfo{volume}{66}},
  \bibinfo{pages}{052406} (\bibinfo{year}{2002}).

\bibitem{Slater1954}
\bibinfo{author}{Slater, J.~C.} \& \bibinfo{author}{Koster, G.~F.}
\newblock \bibinfo{title}{{Simplified {LCAO} Method for the Periodic Potential
  Problem}}.
\newblock \emph{\bibinfo{journal}{Phys. Rev.}} \textbf{\bibinfo{volume}{94}},
  \bibinfo{pages}{1498--1524} (\bibinfo{year}{1954}).

\end{thebibliography}
\bibliographystyle{custom_naturemag}

\appendix
\newpage
\begin{center}
{ \bf \Large Supplementary Material} 
\end{center}

\setcounter{figure}{0}
\renewcommand{\theequation}{\thesection\arabic{equation}}
\renewcommand{\thefigure}{S\arabic{figure}}

\setcounter{equation}{0}
\renewcommand{\thesection}{A}
\setcounter{subsection}{0}

\setcounter{table}{0}
\renewcommand{\thetable}{A\arabic{table}}

\section{Hopping matrices from Slater--Koster parameters}
\label{slaterkoster}

Here we give the explicit Slater--Koster construction underlying Methods Sec.~\ref{subsec:hopping}. 
We use the notation introduced there: $\ket{d_i^G}$ denotes the five-orbital basis centered on site $i$ but expressed in the global cubic frame, $\ket{d_i}$ denotes the corresponding basis in the local octahedral frame, and $\mathcal{R}_{LG}^i$ rotates the global basis into the local basis. 
For a bond $ij$, $\mathcal{T}^{G}_{ij}(l,m,n)$ denotes the Slater--Koster hopping matrix in the global five-orbital basis, with $(l,m,n)$ the direction cosines of the bond from $j$ to $i$.

\subsection{Orbital hopping on the representative $Z$ bond}

The local $t_{2g}$ hopping matrix is obtained by rotating the global Slater--Koster matrix into the local frames at the two sites and projecting to the $t_{2g}$ subspace,
\begin{equation}
\mathcal{T}_{ij}^{t_{2g}}
=
\left[
\mathcal{R}_{LG}^{i}\,
\mathcal{T}_{ij}^{G}(l,m,n)\,
\left(\mathcal{R}_{LG}^{j}\right)^\dagger
\right]_{t_{2g}} .
\label{eq:SM_t2g_hopping_construction}
\end{equation}
For the representative $Z$ bond connecting sites 1 and 4, the direction cosines on the two tetrahedra are
\begin{equation}
(l,m,n)^A_{14}
=
\left(\frac{1}{\sqrt{2}},\frac{1}{\sqrt{2}},0\right),
\qquad
(l,m,n)^B_{14}
=
\left(-\frac{1}{\sqrt{2}},-\frac{1}{\sqrt{2}},0\right).
\end{equation}
Here the superscripts $A$ and $B$ label the two tetrahedra sharing the bond, and the sign difference reflects the opposite orientation of the same bond in the two local tetrahedral environments.
Thus,
\begin{align}
\mathcal{T}_{14,A}^{t_{2g}}
&=
\left[
\mathcal{R}_{LG}^{1}\,
\mathcal{T}_{14,A}^{G}
\left(\frac{1}{\sqrt{2}},\frac{1}{\sqrt{2}},0\right)
\left(\mathcal{R}_{LG}^{4}\right)^\dagger
\right]_{t_{2g}},
\nonumber\\
\mathcal{T}_{14,B}^{t_{2g}}
&=
\left[
\mathcal{R}_{LG}^{1}\,
\mathcal{T}_{14,B}^{G}
\left(-\frac{1}{\sqrt{2}},-\frac{1}{\sqrt{2}},0\right)
\left(\mathcal{R}_{LG}^{4}\right)^\dagger
\right]_{t_{2g}},
\nonumber\\
\mathcal{T}_{41,A}^{t_{2g}}
&=
\left[
\mathcal{R}_{LG}^{4}\,
\mathcal{T}_{41,A}^{G}
\left(-\frac{1}{\sqrt{2}},-\frac{1}{\sqrt{2}},0\right)
\left(\mathcal{R}_{LG}^{1}\right)^\dagger
\right]_{t_{2g}},
\nonumber\\
\mathcal{T}_{41,B}^{t_{2g}}
&=
\left[
\mathcal{R}_{LG}^{4}\,
\mathcal{T}_{41,B}^{G}
\left(\frac{1}{\sqrt{2}},\frac{1}{\sqrt{2}},0\right)
\left(\mathcal{R}_{LG}^{1}\right)^\dagger
\right]_{t_{2g}} .
\end{align}
Evaluating these expressions gives
\begin{equation}
\mathcal{T}_{14,A}^{t_{2g}}
=
\mathcal{T}_{14,B}^{t_{2g}}
=
\begin{pmatrix}
t_1 & t_2 & t_4\\
t_2 & t_1 & t_4\\
-t_4 & -t_4 & t_3
\end{pmatrix},
\qquad
\mathcal{T}_{41,A}^{t_{2g}}
=
\mathcal{T}_{41,B}^{t_{2g}}
=
\begin{pmatrix}
t_1 & t_2 & -t_4\\
t_2 & t_1 & -t_4\\
t_4 & t_4 & t_3
\end{pmatrix}.
\label{eq:SM_Zbond_hopping}
\end{equation}
The equality of the $A$- and $B$-tetrahedron hopping matrices shows that inversion symmetry is preserved in the local basis.

\subsection{Direct hopping amplitudes}

We now express $t_1,t_2,t_3,t_4$ in terms of the Slater--Koster parameters. 
For this purpose, we work in the global five-orbital basis
$\ket{d_i^G}
=
\left(
d^G_{i,YZ},
d^G_{i,XZ},
d^G_{i,XY},
d^G_{i,X^2-Y^2},
d^G_{i,3Z^2-R^2}
\right)^T $.
For the representative $Z$ bond, the corresponding global hopping matrix is
\begin{equation}
\mathcal{T}_{14}^{G}
=
\begin{pmatrix}
\frac{3}{4}(dd\sigma)+\frac{1}{4}(dd\delta) & 0 & -\frac{\sqrt{3}}{4}(dd\sigma)+\frac{\sqrt{3}}{4}(dd\delta) & 0 & 0 \\
0 & dd\pi & 0 & 0 & 0 \\
-\frac{\sqrt{3}}{4}(dd\sigma)+\frac{\sqrt{3}}{4}(dd\delta) & 0 & \frac{1}{4}(dd\sigma)+\frac{3}{4}(dd\delta) & 0 & 0 \\
0 & 0 & 0 & \frac{1}{2}(dd\pi)+\frac{1}{2}(dd\delta) & \frac{1}{2}(dd\pi)-\frac{1}{2}(dd\delta) \\
0 & 0 & 0 & \frac{1}{2}(dd\pi)-\frac{1}{2}(dd\delta) & \frac{1}{2}(dd\pi)+\frac{1}{2}(dd\delta)
\end{pmatrix}
\label{eq:SM_SK_matrix}
\end{equation}

Projecting this matrix onto the local $t_{2g}$ orbitals gives the hopping matrix in Eq.~\eqref{eq:SM_Zbond_hopping}, with
\begin{align}
t_1
&=
\frac{1}{4}
\left[
\big((dd\delta)+4(dd\pi)+3(dd\sigma)\big)\cos 2\varphi
+(dd\delta)-2(dd\pi)-3(dd\sigma)
\right]\cos^2\varphi,
\nonumber\\
t_2
&=
\frac{1}{8}
\left[
-\big((dd\delta)+3(dd\sigma)\big)\sin^2 2\varphi
-4(dd\delta)\cos^2\varphi
+4(dd\pi)\left(\sin^2\varphi+\cos^2 2\varphi\right)
\right],
\nonumber\\
t_3
&=
\frac{1}{32}
\left[
\big((dd\delta)+4(dd\pi)+3(dd\sigma)\big)\cos 4\varphi
+12\big((dd\sigma)-(dd\delta)\big)\cos 2\varphi
+19(dd\delta)-4(dd\pi)+9(dd\sigma)
\right],
\nonumber\\
t_4
&=
\frac{1}{8\sqrt{2}}
\left[
3(dd\delta)-3(dd\sigma)
-\big((dd\delta)+4(dd\pi)+3(dd\sigma)\big)\cos 2\varphi
\right]\sin 2\varphi .
\label{eq:SM_t1234_general}
\end{align}
For ideal octahedra, $\varphi=\arctan(2\sqrt{2})$, and Eq.~\eqref{eq:SM_t1234_general} gives

\begin{eqnarray}
&&\mathcal{T}_{14}^{t_{2g}}
=\frac{1}{162}\times\\\nonumber
&&\begin{pmatrix}
(dd\delta)-23(dd\pi)-24(dd\sigma)
&
-17(dd\delta)+121(dd\pi)-24(dd\sigma)
&
34(dd\delta)+28(dd\pi)-6(dd\sigma)
\\
-17(dd\delta)+121(dd\pi)-24(dd\sigma)
&
(dd\delta)-23(dd\pi)-24(dd\sigma)
&
34(dd\delta)+28(dd\pi)-6(dd\sigma)
\\
-34(dd\delta)-28(dd\pi)+6(dd\sigma)
&
-34(dd\delta)-28(dd\pi)+6(dd\sigma)
&
\frac{289}{2}(dd\delta)-16(dd\pi)+\frac{3}{2}(dd\sigma)
\end{pmatrix}.
\label{eq:SM_direct_hopping_ideal}
\end{eqnarray}

\subsection{Oxygen-mediated hopping amplitudes}
Next we consider the oxygen-mediated hopping. For the oxygen shared by two neighboring octahedra,
    we 
    fix it along 
    the global $(X,Y,Z)$ cubic coordinate. For convenience, we can align the $Z$-axis on the oxygen to $z_1$ and $z_4$ separately when we consider Slater-Koster parameters for the hopping 
between the oxygen ion and the Ru ion, which are given in Table~\ref{tbl:skpd}. We can also find how $p_X$, $p_Y$, $p_Z$ are  expressed in the local reference frames of 
the Ru ions,  $p_{x_1}$, $p_{y_1}$, $p_{z_1}$ on Ru${}_1$  and $p_{x_4}$, $p_{y_4}$, $p_{z_4}$ on Ru${}_4$, respectively:
    \begin{align}
        p_X=&\cos^2\frac{\varphi}{2}p_{x_1}+\frac{\cos\varphi-1}{2}p_{y_1}-\frac{\sin\varphi}{\sqrt{2}}p_{z_1},\nonumber\\
        p_Y=&\frac{\cos\varphi-1}{2}p_{x_1}+\cos^2\frac{\varphi}{2}p_{y_1}-\frac{\sin\varphi}{\sqrt{2}}p_{z_1},\\
        p_Z=&\frac{\sin\varphi}{\sqrt{2}}p_{x_1}+\frac{\sin\varphi}{\sqrt{2}}p_{y_1}+\cos\varphi p_{z_1},\nonumber
    \end{align}
    and
    \begin{align}
        p_X=&\cos^2\frac{\varphi}{2}p_{x_4}+\frac{\cos\varphi-1}{2}p_{y_4}+\frac{\sin\varphi}{\sqrt{2}}p_{z_4},\nonumber\\
        p_Y=&\frac{\cos\varphi-1}{2}p_{x_4}+\cos^2\frac{\varphi}{2}p_{y_4}+\frac{\sin\varphi}{\sqrt{2}}p_{z_4},\\
        p_Z=&-\frac{\sin\varphi}{\sqrt{2}}p_{x_4}-\frac{\sin\varphi}{\sqrt{2}}p_{y_4}+\cos\varphi p_{z_4}.\nonumber
    \end{align}
    This gives us the hopping amplitudes between the global $p$ orbitals on the shared oxygen ion and the local $t_{2g}$ orbitals on two Ru ions in Table~\ref{tbl:skpd}. Suppose the charge transfer energy from d orbital to p orbital is $\Delta_{pd}$, and we can obtain the effective hopping amplitudes between two Ru ions through the shared oxygen as
    \begin{align}
        \begin{pmatrix}
            \dfrac{pd\pi^2}{\Delta_{pd}}\cos^2\varphi & -\dfrac{pd\pi^2}{\Delta_{pd}}\sin^2\varphi & 0\\
            -\dfrac{pd\pi^2}{\Delta_{pd}}\sin^2\varphi & \dfrac{pd\pi^2}{\Delta_{pd}}\cos^2\varphi & 0\\
            0& 0& 0
        \end{pmatrix}.
    \end{align}
    We can absorb these hopping amplitudes into $t_1$ and $t_2$. 
    In the ideal pyrochlore lattice, we have
    
    \begin{table}[h]
        \centering
        \begin{tabular}{|c|c|c|c|}
            \hline
            & $d_{y_1z_1}(d_{y_4z_4})$ & 
            $d_{x_1z_1}(d_{x_4z_4})$ & $d_{x_1y_1}(d_{x_4y_4})$\\\hline
            $p_{x_1}(p_{x_4}) $& $0$ & $pd\pi$ & $0$\\\hline
            $p_{y_1}(p_{y_4})$ & $pd\pi$ & $0$ & $0$\\\hline
            $p_{z_1}(p_{z_4})$ & $0$ & $0$ & $0$\\\hline
        \end{tabular}
        \caption{ The Slater-Koster parameters for hopping between $p$ orbitals to $t_{2g}$ orbitals ($l=0$, $m=0$, $n=1$)}
        \label{tbl:skpd}
    \end{table}

\setcounter{equation}{0}
\renewcommand{\thesection}{B}

\section{Details of basis transformations}
\label{sec:basis_transformations}

For completeness, we summarize the local single-ion basis conventions used in the numerical implementation. 
All states in this section are defined with respect to the local octahedral frame of a given Ru site; the site index is suppressed for notational simplicity.
For one hole in the local $t_{2g}$ shell, or equivalently for a $d^5$ configuration, we use three related bases:\\ the orbital basis
\begin{equation}
\ket{O}
=
\left\{
\ket{X_\uparrow},\ket{X_\downarrow},
\ket{Y_\uparrow},\ket{Y_\downarrow},
\ket{Z_\uparrow},\ket{Z_\downarrow}
\right\},
\end{equation}
where $\ket{X}=d_{yz}$, $\ket{Y}=d_{xz}$, and $\ket{Z}=d_{xy}$;\\
the $LS$ basis
\begin{equation}
\ket{LS}
=
\left\{
\ket{1,\frac12},\ket{1,-\frac12},
\ket{0,\frac12},\ket{0,-\frac12},
\ket{-1,\frac12},\ket{-1,-\frac12}
\right\};
\end{equation}
and the total-angular-momentum basis
\begin{equation}
\ket{J}
=
\left\{
\ket{\frac32,\frac32},
\ket{\frac32,\frac12},
\ket{\frac32,-\frac12},
\ket{\frac32,-\frac32},
\ket{\frac12,\frac12},
\ket{\frac12,-\frac12}
\right\}.
\end{equation}
The orbital angular-momentum states are related to the $t_{2g}$ orbital basis by
\begin{subequations}
\begin{align}
&\ket{0} = \ket{Z},\\
&\ket{1} = -\frac{1}{\sqrt{2}}\left(\ket{X}+i\ket{Y}\right),\\
&\ket{-1} = \frac{1}{\sqrt{2}}\left(\ket{X}-i\ket{Y}\right).
\end{align}
\end{subequations}
In matrix form, the transformation from the angular-momentum basis 
$\{\ket{1},\ket{0},\ket{-1}\}$ to the local orbital basis 
$\{\ket{X},\ket{Y},\ket{Z}\}$ is
\begin{equation}
U_{L\rightarrow O}
=
\begin{pmatrix}
-\frac{1}{\sqrt{2}} & 0 & \frac{1}{\sqrt{2}} \\
-\frac{i}{\sqrt{2}} & 0 & -\frac{i}{\sqrt{2}} \\
0 & 1 & 0
\end{pmatrix}.
\end{equation}
The $LS$ basis is obtained by taking the tensor product of the orbital angular-momentum basis with the spin basis $\{\ket{\uparrow},\ket{\downarrow}\}$. 
Since $U_{L\rightarrow O}$ acts only on the orbital part, the transformation in the full spin--orbital basis is
\begin{equation}
U_{LS\rightarrow O}=U_{L\rightarrow O}\otimes \mathbb{1}_{2}.
\end{equation}
The transformation from the $LS$ basis to the total-angular-momentum basis is fixed by Clebsch-Gordan coefficients:
\begin{equation}
\left(
\begin{array}{l}
\ket{\frac32,\frac32}\\
\ket{\frac32,\frac12}\\
\ket{\frac32,-\frac12}\\
\ket{\frac32,-\frac32}\\
\ket{\frac12,\frac12}\\
\ket{\frac12,-\frac12}
\end{array}
\right)
=
U_{LS\rightarrow J}
\left(
\begin{array}{l}
\ket{1,\frac12}\\
\ket{1,-\frac12}\\
\ket{0,\frac12}\\
\ket{0,-\frac12}\\
\ket{-1,\frac12}\\
\ket{-1,-\frac12}
\end{array}
\right),
\end{equation}
with
\begin{equation}
U_{LS\rightarrow J}
=
\begin{pmatrix}
1 & 0 & 0 & 0 & 0 & 0 \\
0 & \frac{1}{\sqrt{3}} & \sqrt{\frac{2}{3}} & 0 & 0 & 0 \\
0 & 0 & 0 & \sqrt{\frac{2}{3}} & \frac{1}{\sqrt{3}} & 0 \\
0 & 0 & 0 & 0 & 0 & 1 \\
0 & \sqrt{\frac{2}{3}} & -\frac{1}{\sqrt{3}} & 0 & 0 & 0 \\
0 & 0 & 0 & \frac{1}{\sqrt{3}} & -\sqrt{\frac{2}{3}} & 0
\end{pmatrix}.
\end{equation}
These conventions fix the phases used in constructing the local spin--orbit states and the numerical projection matrices.

\setcounter{equation}{0}
\setcounter{subsection}{0}
\renewcommand{\thesection}{C}

\section{Symmetry-related bond Hamiltonians}
\label{sec:symmetry_bonds}

In Methods, the projected exchange Hamiltonian is constructed explicitly for the representative $Z$ bond connecting sites 1 and 4. 
Here we describe how the remaining nearest-neighbor bond Hamiltonians are generated from this representative bond by symmetry.

Starting from the projected Hamiltonian $\mathcal{H}^{Z}_{14}$ on the representative $Z$ bond, the Hamiltonians on the symmetry-related bonds are obtained by applying rotations of the tetrahedral point group $T_d$. 
These rotations map one local bond environment to another. 
Since the singlet is invariant, their only nontrivial action in the low-energy Hilbert space is to permute the Cartesian triplet components.

For example, a $C_3$ rotation about the $[111]$ axis maps the local environment of the representative $Z$ bond $(1,4)$ to that of the $X$ bond $(1,2)$. 
We denote the basis after this rotation by $\ket{\widetilde{\tau}_\alpha}$ and choose the convention
\begin{equation}
\ket{\tau_\alpha}
=
P_{111}\ket{\widetilde{\tau}_\alpha}.
\end{equation}
With this convention, the projected Hamiltonian in the rotated basis is related to the representative $Z$-bond Hamiltonian by
\begin{equation}
\begin{aligned}
&
\bra{\widetilde{\tau}_{1\alpha} \widetilde{\tau}_{2\beta}}
\mathcal{H}^{X}_{12}
\ket{\widetilde{\tau}_{1\gamma} \widetilde{\tau}_{2\delta}}
\\
&\qquad =
\bra{\tau_{1\alpha} \tau_{2\beta}}
\left(P_{111}\otimes P_{111}\right)
\mathcal{H}^{Z}_{14}
\left(P^{-1}_{111}\otimes P^{-1}_{111}\right)
\ket{\tau_{1\gamma} \tau_{2\delta}}.
\end{aligned}
\label{eq:SM_HX_from_HZ_matrix_elements}
\end{equation}
Equivalently, at the matrix level,
\begin{equation}
\mathcal{H}^{X}_{12}
=
\left(P_{111}\otimes P_{111}\right)
\mathcal{H}^{Z}_{14}
\left(P^{-1}_{111}\otimes P^{-1}_{111}\right),
\label{eq:SM_HX_from_HZ}
\end{equation}
with the site labels relabeled from $(1,4)$ to $(1,2)$.

The permutation matrices used in the construction are
\begin{equation}
P_{111}
=
\begin{pmatrix}
 1 & 0 & 0 & 0 \\
 0 & 0 & 0 & -1 \\
 0 & 1 & 0 & 0 \\
 0 & 0 & -1 & 0
\end{pmatrix},
\qquad
P_{1\bar{1}\bar{1}}
=
\begin{pmatrix}
 1 & 0 & 0 & 0 \\
 0 & 0 & 0 & 1 \\
 0 & -1 & 0 & 0 \\
 0 & 0 & -1 & 0
\end{pmatrix}.
\label{eq:SM_permutation_matrices}
\end{equation}
In each matrix, the upper-left entry acts on the singlet, which is invariant, while the lower $3\times3$ block acts on the Cartesian triplet components.

One convenient sequence of symmetry operations generating the six nearest-neighbor bond Hamiltonians on a tetrahedron is
\begin{equation}
\begin{gathered}
\mathcal{H}^{Z}_{14}
\xrightarrow{C_3[111]}
\mathcal{H}^{X}_{12}
\xrightarrow{C_3[111]}
\mathcal{H}^{Y}_{13},
\\
\mathcal{H}^{X}_{12}
\xrightarrow{C_3[1\bar{1}\bar{1}]}
\mathcal{H}^{Y}_{42}
\xrightarrow{C_3[1\bar{1}\bar{1}]}
\mathcal{H}^{Z}_{32},
\\
\mathcal{H}^{Z}_{14}
\xrightarrow{C_3[1\bar{1}\bar{1}]}
\mathcal{H}^{X}_{43}.
\end{gathered}
\label{eq:SM_bond_generation}
\end{equation}
For example, the first relation in Eq.~\eqref{eq:SM_bond_generation} is implemented by Eq.~\eqref{eq:SM_HX_from_HZ}. 
Similarly, the transformation 
$\mathcal{H}^{X}_{12}\xrightarrow{C_3[1\bar{1}\bar{1}]}\mathcal{H}^{Y}_{42}$ is implemented using $P_{1\bar{1}\bar{1}}$,
\begin{equation}
\mathcal{H}^{Y}_{42}
=
\left(P_{1\bar{1}\bar{1}}\otimes P_{1\bar{1}\bar{1}}\right)
\mathcal{H}^{X}_{12}
\left(P^{-1}_{1\bar{1}\bar{1}}\otimes P^{-1}_{1\bar{1}\bar{1}}\right), 
\end{equation}
and the remaining relations are obtained analogously. 
Together, the six matrices
\begin{equation}
\left\{
\mathcal{H}^{Z}_{14},
\mathcal{H}^{X}_{12},
\mathcal{H}^{Y}_{13},
\mathcal{H}^{Z}_{32},
\mathcal{H}^{X}_{43},
\mathcal{H}^{Y}_{42}
\right\}
\end{equation}
form the nearest-neighbor building blocks used to assemble the full exchange Hamiltonian on the pyrochlore lattice.

\setcounter{equation}{0}
\setcounter{subsection}{0}
\renewcommand{\thesection}{D}
\section*{D. Harmonic Approximation of the Quartic Hamiltonian}\label{app:quadham}

To analyze the instability toward triplon condensation and the onset of magnetic ordering, we study small fluctuations around the singlet-condensed phase, where the triplon operators describe low-density triplet excitations above the nonmagnetic singlet background. The local constraint,
\begin{equation}
    n_s+\sum_{\alpha=x,y,z} n_{T^\alpha}=1,
\end{equation}
is enforced within a Gutzwiller approximation by replacing the singlet operators according to
\begin{equation}
    s,s^\dagger
    \rightarrow
    \sqrt{1-\sum_\alpha T^{\alpha\dagger} T^{\alpha}}.
\end{equation}
Expanding around the singlet-condensed state and retaining terms up to quadratic order in the triplon operators yields the harmonic triplon Hamiltonian.

The quartic superexchange Hamiltonian obtained from the second order perturbation theory is 
\begin{equation}\label{H4SM}
    \mathcal{H}^{(4)}_{\mathrm{eff}}=\sum_{\phi,\phi'} \ket{\phi}\braket{\phi|\mathcal{H}_{\text{eff}}|\phi'}\bra{\phi'} = \sum_{\braket{ij},\alpha\alpha'\beta\beta'}J^{\alpha\alpha'\beta\beta'}_{ij}\, \uptau_{i\alpha}^\dagger\uptau_{j\alpha'}^\dagger\uptau_{i\beta}\uptau_{j\beta'} + h.c.,
\end{equation}
where $J^{\alpha\alpha'\beta\beta'}_{ij}$ (with
$\alpha,\alpha',\beta,\beta'=0,x,y,z$) are the effective interaction amplitudes in the two-site singlet-triplet basis
$\ket{\phi}=\ket{\tau_{i\alpha}\rangle\otimes|\tau_{j\alpha'}}$.

 After substituting the Gutzwiller expansion for the singlet operators and retaining terms up to quadratic order in the triplon fields, only quartic operator structures containing at least two singlet operators contribute within the harmonic approximation. Among the 16 possible quartic terms appearing in Eq.~(\ref{H4SM}), the following seven generate constant, onsite, hopping, or pairing contributions:
\begin{enumerate}
    \item $s_i^\dagger s_j^\dagger s_i s_j  \sim 1 - \sum_\alpha (T_{i\alpha}^\dagger T_{i\alpha}+ T_{j\alpha}^\dagger T_{j\alpha})+ \mathcal{O}(T^\dagger T)^2$ 
    \item $s_i^\dagger s_j^\dagger T_{i\alpha} T_{j\alpha'} \sim  T_{i\alpha} T_{j\alpha'} +\mathcal{O}(TT)^2$
    \item $s_i^\dagger T_{j\alpha}^\dagger s_{i} T_{j\alpha'} \sim T_{j\alpha}^\dagger T_{j\alpha'}+ \mathcal{O}(T^\dagger T)^2$
    \item $s_i^\dagger T_{j\alpha}^\dagger T_{i\alpha'}s_{j} \sim T_{j\alpha}^\dagger T_{i\alpha'}+ \mathcal{O}(T^\dagger T)^2$
    \item $ T_{i\alpha}^\dagger s_j^\dagger T_{i\alpha'}s_{j} \sim T_{i\alpha}^\dagger T_{i\alpha'} + \mathcal{O}(T^\dagger T)^2$
    \item $ T_{i\alpha}^\dagger s_j^\dagger s_{i} T_{j\alpha'} \sim T_{i\alpha}^\dagger T_{j\alpha'}+\mathcal{O}(T^\dagger T)^2$
    \item $T^\dagger_{i\alpha} T^\dagger_{j\alpha'} s_i s_j \sim T^\dagger_{i\alpha} T^\dagger_{j\alpha'} + \mathcal{O}(T^\dagger T^\dagger)^2$
\end{enumerate}
Thus, the quartic Hamiltonian reduces to a quadratic form containing normal and anomalous triplon terms,
\begin{equation}\label{H4Decoupled}
    \mathcal{H}^{(2)}_{\mathrm{eff}}
    =
    \sum_{\langle ij\rangle,\alpha\alpha'}
    \left[
    A^{\alpha\alpha'}_{ij}\,
    T^\dagger_{i\alpha}T_{j\alpha'}
    +
    \frac{1}{2}
    \left(
    B^{\alpha\alpha'}_{ij}\,
    T^\dagger_{i\alpha}T^\dagger_{j\alpha'}
    +
    \mathrm{h.c.}
    \right)
    \right],
\end{equation}
where $A^{\alpha\alpha'}_{ij}$ and $B^{\alpha\alpha'}_{ij}$ are obtained from the corresponding coefficients of the full quartic Hamiltonian.

Since the harmonic Hamiltonian is quadratic in the triplon operators and preserves lattice translational symmetry, it is convenient to work in momentum space, where different crystal momenta decouple. The resulting bosonic Bogoliubov-de Gennes Hamiltonian, which determines the triplon dispersion and allows one to identify possible instabilities of the singlet phase toward triplon condensation,
can be written  as
\begin{equation}
    \mathcal{H}^{(2)}_{\mathrm{eff}}
    =
    \sum_{\mathbf{k}}
    \Psi^\dagger_{\mathbf{k}}
    \begin{pmatrix}
        \mathcal{A}_{\mathbf{k}} & \mathcal{B}_{\mathbf{k}} \\
        \mathcal{B}_{\mathbf{k}}^\dagger & \mathcal{A}_{-\mathbf{k}}^{T}
    \end{pmatrix}
    \Psi_{\mathbf{k}},
\end{equation}
where
\begin{equation}
    \Psi_{\mathbf{k}}
    =
    \begin{pmatrix}
        \mathbf{T}_{\mathbf{k}} \\
        \mathbf{T}_{-\mathbf{k}}^\dagger
    \end{pmatrix},
    \quad
    \mathbf{T}_{\mathbf{k}}
    =
    \left(
    T_{1x,\mathbf{k}},
    T_{1y,\mathbf{k}},
    T_{1z,\mathbf{k}},
    \ldots,
    T_{4x,\mathbf{k}},
    T_{4y,\mathbf{k}},
    T_{4z,\mathbf{k}}
    \right)^T.
\end{equation}
The matrices $\mathcal{A}_{\mathbf{k}}$ and $\mathcal{B}_{\mathbf{k}}$ are $12\times12$ matrices describing normal hopping and anomalous pairing processes between triplons on the four pyrochlore sublattices.

The normal block $\mathcal{A}_{\mathbf{k}}$ has the general sublattice structure
\begin{equation}
    \mathcal{A}_{\mathbf{k}}
    =
    \begin{pmatrix}
        A_{11} & A_{12}^{X} & A_{13}^{Y} & A_{14}^{Z} \\
        A_{21}^{X} & A_{22} & A_{23}^{Z} & A_{24}^{Y} \\
        A_{31}^{Y} & A_{32}^{Z} & A_{33} & A_{34}^{X} \\
        A_{41}^{Z} & A_{42}^{Y} & A_{43}^{X} & A_{44}
    \end{pmatrix},
\end{equation}
where each entry is itself a $3\times3$ matrix in the  triplon flavor space. The subscripts denote the pyrochlore sublattices, the superscripts label the corresponding bond type, and the $\mathbf{k}$ dependence of all entries is implicit.

As an illustrative example, consider the contribution from the representative $Z$ bond connecting sites $(1,4)$. In momentum space, the corresponding matrix elements generate normal triplon terms of the form
\begin{align}
\mathcal{H}^{Z}_{14,\mathbf{k}}&=\mathcal{H}_{\mathrm{eff}}^{(2)}+\mathcal{H}_{\mathrm{SOC},1}+\mathcal{H}_{\mathrm{SOC},4}
=
\begin{pmatrix}
     \left(\mathcal{H}^{Z}_{14,\mathbf{k}}\right)_{\text{onsite}/4}
     &
     \left(\mathcal{H}^{Z}_{14,\mathbf{k}}\right)_{4\leftarrow 1}
     \\
     \left(\mathcal{H}^{Z}_{14,\mathbf{k}}\right)_{1\leftarrow 4}
     &
     \left(\mathcal{H}^{Z}_{14,\mathbf{k}}\right)_{\text{onsite}/1}
\end{pmatrix}
\nonumber \\\vspace{3pt}
&=
\left(
    \begin{array}{c|ccc|ccc}
        & \ket{s_1T_{4x}} & \ket{s_1T_{4y}} & \ket{s_1T_{4z}}
        & \ket{T_{1x}s_4} & \ket{T_{1y}s_4} & \ket{T_{1z}s_4} \\
        \hline
        \bra{s_1T_{4x}} & \ddots & & & \ddots  & & \\
        \bra{s_1T_{4y}} & & \boxed{\#T^\dagger_{4\alpha,\mathbf{k}}T_{4\alpha',\mathbf{k}}} & & & \boxed{\#T^\dagger_{4\alpha,\mathbf{k}}T_{1\alpha',\mathbf{k}}} & \\
        \bra{s_1T_{4z}} & & & \ddots  & & & \ddots \\
        \hline 
        \bra{T_{1x}s_4} & \ddots & & & \ddots & & \\
        \bra{T_{1y}s_4} & & \boxed{\#T^\dagger_{1\alpha,\mathbf{k}}T_{4\alpha',\mathbf{k}}}  & & & \boxed{\#T^\dagger_{1\alpha,\mathbf{k}}T_{1\alpha',\mathbf{k}}} &  \\
        \bra{T_{1z}s_4} & & & \ddots & & & \ddots 
    \end{array}
    \right).
\end{align}
 The diagonal blocks generate onsite contributions of the form
$T^\dagger_{i\alpha,\mathbf{k}}T_{i\alpha',\mathbf{k}}$,
while the off-diagonal blocks describe intersite hopping processes
$T^\dagger_{i\alpha,\mathbf{k}}T_{j\alpha',\mathbf{k}}$.
Each block is a matrix in the triplon flavor space $\alpha=x,y,z$.

The anomalous block $\mathcal{B}_{\mathbf{k}}$ is constructed in the same way from pair-creation terms of the form
$T^\dagger_{i\alpha,\mathbf{k}}T^\dagger_{j\alpha',-\mathbf{k}}$.
These terms originate from matrix elements connecting the singlet-singlet state to two-triplon states,
$\ket{T_i^\alpha T_j^{\alpha'}}\bra{s_i s_j},$
in the quartic Hamiltonian. Since there are no onsite pair-creation terms of the form
$T^\dagger_{i\alpha,\mathbf{k}}T^\dagger_{i\alpha',-\mathbf{k}}$, the diagonal sublattice blocks vanish. Thus,
\begin{equation}
    \mathcal{B}_{\mathbf{k}}
    =
    \begin{pmatrix}
        \mathbf{0} & B_{12}^{X} & B_{13}^{Y} & B_{14}^{Z} \\
        B_{21}^{X} & \mathbf{0} & B_{23}^{Z} & B_{24}^{Y} \\
        B_{31}^{Y} & B_{32}^{Z} & \mathbf{0} & B_{34}^{X} \\
        B_{41}^{Z} & B_{42}^{Y} & B_{43}^{X} & \mathbf{0}
    \end{pmatrix},
\end{equation}
where each entry is again a $3\times3$ matrix in triplon flavor space, and the $\mathbf{k}$ dependence is implicit.

Once we construct the quadratic Hamiltonian, it is diagonalized in momentum space via the Fourier transform
\begin{equation}\label{eqn:fourier}
    \mathbf{T}^\dagger_{\mu,\mathbf{k}} = \frac{1}{\sqrt{N}}\sum_{\mathbf{r}} e^{i \mathbf{k}\cdot\mathbf{r}} \,\mathbf{T}^\dagger_{\mu,\mathbf{r}}, \quad  \mathbf{T}_{\mu,\mathbf{k}} = \frac{1}{\sqrt{N}}\sum_{\mathbf{r}} e^{-i \mathbf{k}\cdot\mathbf{r}} \,\mathbf{T}_{\mu,\mathbf{r}},
\end{equation}
where $\mu$ is the sublattice index. The quadratic Hamiltonian can be expressed in the triplon basis $\Gamma_\mathbf{k} = (\mathbf{T}_{1,\mathbf{k}},\mathbf{T}_{2,\mathbf{k}},\mathbf{T}_{3,\mathbf{k}},\mathbf{T}_{4,\mathbf{k}},\mathbf{T}^\dagger_{1,\mathbf{-k}}, \mathbf{T}^\dagger_{2,\mathbf{-k}},\mathbf{T}^\dagger_{3,\mathbf{-k}},\mathbf{T}^\dagger_{4,\mathbf{-k}})^T$ where each triplon comes in three flavors $\mathbf{T}_{\mu,\mathbf{k}}=(T_{\mu,\mathbf{k}}^x,T_{\mu,\mathbf{k}}^y,T_{\mu,\mathbf{k}}^z)$. Thus, in the harmonic approximation, the Hamiltonian can be expressed in matrix form as
\begin{equation}
    \mathcal{H}^{(2)} = \sum_\mathbf{k} \Gamma_\mathbf{k}^\dagger H_\mathbf{k} \Gamma_\mathbf{k}
\end{equation}
$H_\mathbf{k}$ can be diagonalized by a bosonic Bogoliubov transformation $\Gamma_\mathbf{k} = \mathcal{S}_\mathbf{k} \tilde{\Gamma}_\mathbf{k}$ such that the bosonic commutation is enforced by the metric tensor $g\equiv\text{diag}(\mathbb{1},-\mathbb{1})$ through $\mathcal{S}_\mathbf{k}^\dagger g \mathcal{S}_\mathbf{k} = g$. Thus, the diagonalized Hamiltonian 
\begin{equation}
    \mathcal{H}^{(2)} = \sum_\mathbf{k}\sum_{\mu=1}^{12} \omega_{\mu,\mathbf{k}} \bigg( \tilde{T}^\dagger_{\mu,\mathbf{k}} \tilde{T}_{\mu,\mathbf{k}} + \frac{1}{2}\bigg)
\end{equation}
gives 12 physical bands characterized by bosonic modes, $\tilde{\Gamma}_\mathbf{k}=(\tilde{T}_{1,\mathbf{k}},\tilde{T}_{2,\mathbf{k}},\tilde{T}_{3,\mathbf{k}}\ldots, \tilde{T}_{12,\mathbf{k}},\tilde{T}^{\dagger}_{1,-\mathbf{k}},\ldots,\tilde{T}^{\dagger}_{12,-\mathbf{k}})^T$, each with frequency $\omega_{\mu,\mathbf{k}}$.

\end{document}